\documentclass[fleqn,usenatbib]{mnras}

\usepackage{newtxtext,newtxmath}
\usepackage[T1]{fontenc}

\DeclareRobustCommand{\VAN}[3]{#2}
\let\VANthebibliography\thebibliography
\def\thebibliography{\DeclareRobustCommand{\VAN}[3]{##3}\VANthebibliography}

\usepackage{graphicx}	% Including figure files
\usepackage{amsmath}	% Advanced maths commands
\usepackage{color}
\usepackage{tikz}

\usepackage{array}
\usepackage{multirow}

\newcommand\MyBox[2]{
  \fbox{\lower0.75cm
    \vbox to 1.7cm{\vfil
      \hbox to 2.5cm{\hfil\parbox{2.0cm}{#1\\#2}\hfil}
      \vfil}%
  }%
}

\definecolor{linkcolor}{rgb}{0,0,0.25}
\hypersetup{
colorlinks=true,        % false: boxed links; true: colored links
linkcolor=blue,    % color of internal links
citecolor=blue,    % color of links to bibliography
filecolor=blue,    % color of file links
urlcolor=blue,      % color of external links
draft=False,
}

\makeatletter
\renewcommand{\@printed}{}
\makeatother

\newcommand{\figurename}{Figure}
\newcommand{\tablename}{Table}
\newcommand{\eqnname}{Equation}
\newcommand{\secname}{Section}

\definecolor{darkgreen}{rgb}{0.0, 0.7, 0.0}

\definecolor{darkblue}{rgb}{0.0, 0., 0.7}

\newcommand{\gaia}{\emph{Gaia}}
\newcommand{\tess}{\emph{TESS}}
\newcommand{\kepler}{\emph{Kepler}}
\newcommand{\corot}{\emph{CoRoT}}

\renewcommand{\vec}[1]{\ensuremath{\mathbf{#1}}}
\newcommand{\teff}{\ensuremath{T_\mathrm{eff}}}
\newcommand{\logg}{\ensuremath{\log g}}
\newcommand{\xh}[1]{\ensuremath{[\mathrm{#1/H}]}}
\newcommand{\xfe}[1]{\ensuremath{[\mathrm{#1/Fe}]}}
\newcommand{\cn}{\ensuremath{[\mathrm{C/N}]}}
\newcommand{\alpham}{\ensuremath{[\alpha/\mathrm{M}]}}

\newcommand{\um}{\ensuremath{\mu m}}

\newcommand{\numax}{\ensuremath{\nu_{\mathrm{max}}}}
\newcommand{\Dnu}{\ensuremath{\Delta \nu}}
\newcommand{\uHz}{\ensuremath{\mu\mathrm{Hz}}}

\newcommand{\approptoinn}[2]{\mathrel{\vcenter{
  \offinterlineskip\halign{\hfil$##$\cr
    #1\propto\cr\noalign{\kern2pt}#1\sim\cr\noalign{\kern-2pt}}}}}

\newcommand{\appropto}{\mathpalette\approptoinn\relax}
\title[Spectroscopic ages with an encoder-decoder]{A variational encoder-decoder approach to precise spectroscopic age estimation for large Galactic surveys}%: Spectroscopic ages using a variational encoder-decoder}
\author[Leung et al.]{
Henry W. Leung$^{1}$\thanks{E-mail: henrysky.leung@utoronto.ca},
Jo Bovy$^{1,2}$,
J.~Ted Mackereth$^{1,2,3}$, and 
Andrea Miglio$^{4}$
\newauthor
\\
$^{1}$David A. Dunlap Department of Astronomy and Astrophysics, University of Toronto, 50 St. George Street, Toronto, Ontario, M5S 3H4, Canada\\
$^{2}$Dunlap Institute for Astronomy and Astrophysics, University of Toronto, 50 St. George Street, Toronto, Ontario, M5S 3H4, Canada\\
$^{3}$Canadian Institute for Theoretical Astrophysics, University of Toronto, 60 St George Street, Toronto, ON M5S 3H8, Canada\\
$^{4}$Dipartimento di Fisica e Astronomia, Universit{\'a} degli Studi di Bologna, Via Gobetti 93/2, I-40129 Bologna, Italy
}

\date{Accepted XXX. Received YYY; in original form ZZZ}

\pubyear{2023}

\begin{document}
\label{firstpage}
\pagerange{\pageref{firstpage}--\pageref{lastpage}}
\maketitle

\begin{abstract}
Constraints on the formation and evolution of the Milky Way Galaxy require multi-dimensional measurements of kinematics, abundances, and ages for a large population of stars. Ages for luminous giants, which can be seen to large distances, are an essential component of studies of the Milky Way, but they are traditionally very difficult to estimate precisely for a large dataset and often require careful analysis on a star-by-star basis in asteroseismology. Because spectra are easier to obtain for large samples, being able to determine precise ages from spectra allows for large age samples to be constructed, but spectroscopic ages are often imprecise and contaminated by abundance correlations.  
Here we present an application of a variational encoder-decoder on cross-domain astronomical data to solve these issues. The model is trained on pairs of observations from APOGEE and \kepler\ of the same star in order to reduce the dimensionality of the APOGEE spectra in a latent space while removing abundance information. The low dimensional latent representation of these spectra can then be trained to predict age with just $\sim 1,000$ precise seismic ages. We demonstrate that this model produces more precise spectroscopic ages ($\sim 22\%$ overall, $\sim 11\%$ for red-clump stars) than previous data-driven spectroscopic ages while being less contaminated by abundance information (in particular, our ages do not depend on \alpham). We create a public age catalog for the APOGEE DR17 data set and use it to map the age distribution and the age-\xh{Fe}-\alpham\ distribution across the radial range of the Galactic disk.
\end{abstract}


%%%%%%%%%%%%%%%%%%%%%%%%%%%%%%%%%%%%%%%%%%%%%%%%%%

%%%%%%%%%%%%%%%%% BODY OF PAPER %%%%%%%%%%%%%%%%%%

\section{Introduction}
Stars provide an important window into our Galaxy's evolutionary history as all major events that occurred in the past leave imprints in the chemical abundances and kinematics of stars \citep{2002ARA&A..40..487F}. To improve our understanding of the formation history of the Milky Way and explore the evolution and chemo-dynamical structure of the Galaxy as a whole, we need to measure abundances and kinematics of stars as functions of age for stellar samples covering a large volume of our Galaxy from the bulge and the disk to the stellar halo \citep{2013A&ARv..21...61R, 2016ARA&A..54..529B}. Age, chemical abundances and kinematics are interconnected in complex ways (e.g., \citealt{1993A&A...275..101E,2013A&A...560A.109H,2019MNRAS.489..176M, 2019ApJ...883..177N, 2019MNRAS.490.4740B}) and information on their distributions far away from the solar neighbourhood remains scant. To observe a large volume of stars for galactic archaeology purposes, low-mass giants are of particular interest, because they are common throughout the Galaxy, live relatively long and stable lives, and they are intrinsically luminous allowing them to be observed to large distances even in regions with high extinction such as the Galactic bulge.

Modern spectroscopic surveys such as SDSS-IV's APOGEE \citep{2017AJ....154...94M, 2017AJ....154...28B}, GALAH \citep{2015MNRAS.449.2604D}, the ongoing SDSS-V's Milky Way Mapper (MWM; \citealt{2017arXiv171103234K}), and \gaia\ \citep{2016A&A...595A...1G} provide accurate measurements of basic stellar parameters like \teff, elemental abundances, and kinematics. These surveys, however, do not directly provide accurate stellar age measurements for low-mass giants, because stellar ages are not a directly observable quantity. Unlike sub-giants, for which age can be measured fairly accurately with basic stellar parameters and using isochrones \citep{2013A&A...560A.109H,2017ApJS..232....2X, 2022Natur.603..599X}, stellar ages for giants are intrinsically difficult to measure, because giant evolutionary tracks are crowded together compared to sub-giant isochrones, age and metallicity are to some extent degenerate observationally, and stellar evolutionary models have large uncertainties. While age correlates with kinematics and abundances, individual stellar ages cannot simply be inferred accurately from stellar kinematics (e.g., \citealt{2018ApJ...867...31B}), abundance, or kinematics-abundance alone. Even if ages could be inferred in this way, to investigate the relation between age, abundance, and kinematics, we cannot rely on pre-determined relations in this space.

Ages for giant stars can be obtained from determinations of their mass, because the age of a giant is almost directly given by its mass-dependent main-sequence (MS) lifetime. Because of the steep dependence of the MS lifetime on mass ($\tau \appropto M^{-3.5}$), small uncertainties in mass determinations get amplified strongly and giant ages typically have large uncertainties. Spectroscopically, masses for giants can be determined using proxies such as the \cn\ ratio (e.g., \citealt{2015MNRAS.453.1855M, 2016MNRAS.456.3655M}), which is partially set by the mass-dependent dredge-up process. Alternatively, accurate masses for giants can be obtained with careful analysis on a star-by-star basis \citep{2020A&A...642A.226A} using asteroseismic observations, which can determine masses from the properties of stochastically-driven oscillations that can be observed by space telescopes such as \corot\ \citep{2009A&A...506..411A}, \kepler\ \citep{2010Sci...327..977B} and \tess\ \citep{2015JATIS...1a4003R}. Using stellar masses in combination with spectroscopically determined parameters  like \teff\ and \xh{Fe}, we can derive stellar age using stellar models (e.g., \citealt{2014MNRAS.445.2758R}). The APOKASC project \citep{2014ApJS..215...19P, 2018ApJS..239...32P} is an example of this, combining \kepler\ seismic observation with APOGEE spectroscopic parameters to derive stellar ages with uncertainties of $\approx 30\%$. APOKASC has opened up the possibility of doing galactic archaeology with ages for thousands of giants, but it is limited to the relatively small \kepler\ field. While TESS' sky coverage is much bigger, its shorter observation span means that it is difficult to use TESS to determine asteroseismic ages for high-luminosity giants with their long oscillation time scales. 

Spectroscopic surveys such as APOGEE or the ongoing SDSS-V Milky Way Mapper (MWM) have all-sky coverage and obtain spectra for $\approx$ 1 million (for APOGEE) to 5 million (MWM) stars. Recent advances in machine learning methodology and algorithms allow us to do transfer learning, using which we can transfer knowledge obtained in one domain to another by training on pairs of data from the two domains. In the context of ages, this means that we can transfer age knowledge from the asteroseismic realm to the spectroscopic realm by training on stars with observations in both realms, without requiring any prior knowledge on how to map stellar spectra to ages. The APOKASC catalog provides such a training set and it has been used to determine spectroscopic ages using APOGEE (e.g. \citealt{2016ApJ...823..114N,2019MNRAS.489..176M, 2021MNRAS.503.2814C}), LAMOST (e.g., \citealt{2019ApJS..245...34X}), and other surveys. This has allowed for millions of data-driven spectroscopic ages to be determined. As \citet{2019ApJS..245...34X} demonstrates, these spectroscopic ages provide great scientific value even if their precision falls short of what is ideally required.

Spectroscopic ages obtained through transfer-learning from asteroseismic data currently suffer from a series of limitations. Firstly, the overlap between spectroscopic surveys and asteroseismic surveys is small with $\mathcal{O}(10^4)$ stars. This is a relatively small amount of data to train modern machine-learning methods (e.g., neural networks; \citealt{2019MNRAS.489..176M, 2021MNRAS.503.2814C}). However, many more pairs of, e.g., APOGEE/Kepler observations exist than we have asteroseismic ages for and these could in principle be used to improve the information transfer between the domains. Secondly, spectra contain information on abundances that are highly correlated with age (e.g., the alpha enhancement $\xfe{\alpha}$) and current methods provide no guarantee that the spectroscopically-determined age is not solely or largely coming from age-abundance correlations present in the training sample rather than true spectral age information. This makes any inference of the age-abundance-kinematics correlations using current spectroscopic ages suspect. 

In this paper, we present a novel transfer-learning method for determining spectroscopic ages for giants that solves these issues and furthermore allows for spectroscopic ages to be determined in the future using other small, high-quality samples of stellar ages. We do this by splitting the spectroscopic-age determination task in two parts: (i) extracting the age information from stellar spectra while  discarding abundance information and (ii) mapping the extracted age information to age using a small sample of accurate stellar ages. We achieve (i) by using a variational encoder-decoder neural network to map high-resolution spectra to asteroseismic power spectra using a small latent-space bottleneck connecting the two. Because asteroseismic power spectra contain information on mass and radius but not abundance, this effectively extracts age information from the stellar spectra while discarding abundance information. In step (ii) we then train a simpler machine-learning method with fewer free parameters to map the latent space to age using the small sample of high-quality ages. Predictions for new spectroscopic ages are obtained by encoding the spectrum in the latent space and then mapping its location in latent space to age. We then apply this method to the APOGEE data and map the age distribution and age--abundance correlations across the Galactic disk.

This paper is organized as follows. We given an overview of the relevant deep-learning methodology in \secname\ \ref{sec:methodology}. \secname\ \ref{sec:model} describes the actual machine-learning methods that we use: the encoder-decoder used in step (i) of the algorithm in \secname\,\ref{subsec:endecoder} and a modified version of the random forest method used in step (ii) in \secname\,\ref{subsec:rf}. We then discuss the data from APOGEE, \kepler, and APOKASC that we use in \secname\ \ref{sec:data}. We give details on the training and validation steps of our algorithm in \secname\ \ref{sec:traintest} and then describe the results in \secname\ \ref{sec:result}: in \secname\,\ref{subsec:psd_recon} we show how well we can reconstruct asteroseismic power spectra from high-resolution spectra using the encoder-decoder network, we discuss the latent space representation of the spectra in \secname\,\ref{subsec:latent}, we discuss the derived seismic parameters, ages, and evolutionary state classifications in \secname\,\ref{subsec:realresult}. The detail of applying our method to generate APOGEE age catalog for APOGEE DR17 is given in \secname\,\ref{sec:dr17_catalog} and spatial age-abundance trends in the Galactic disk are shown in \secname\,\ref{subsec:agetrend}. We discuss the implications of our results and possible future applications in \secname\,\ref{sec:discussion} and then conclude  in \secname\,\ref{sec:conclusion}.

\section{Deep Learning Methodology} \label{sec:methodology}

In this section, we provide an overview of the deep-learning methodology that we use in this work: (variational) auto-encoders in \secname\,\ref{subsec:autoencoder}, encoder-decoder networks in \secname\,\ref{subsec:endecode}, and we end with a discussion of the rationale behind why we choose to use an encoder-decoder in this work in \secname\,\ref{subsec:rationale}.

\subsection{Variational auto-encoders} \label{subsec:autoencoder}

Deep learning using artificial neural networks is a versatile and flexible machine-learning method and it has been applied to problems in both supervised and unsupervised learning with data ranging from two-dimensional images, voice recordings, video, and sequences of words \citep{2015Natur.521..436L}. Auto-encoders are a specific type of neural network primarily developed to learn efficient, low-dimensional representations of input data that allow the input to be faithfully reproduced. As such, auto-encoders are essentially a non-linear version of Principal Component Analysis (PCA), that is an auto-encoder network without any non-linearity (e.g., not using non-linear activation functions such as Rectified linear units; \citealt{10.5555/3104322.3104425}) using a squared-error loss function are equivalent to traditional PCA \citep{cdi_webofscience_primary_A1991EX55000008}. The basic idea of using an auto-encoder is having a network that takes an input image, compresses it to a low-dimensional middle layer (the \emph{latent space}) in a sequence of layers (\emph{the encoder}), and finally decompresses this low-dimensional representation to reconstruct the input image (the \emph{decoder}). The low-dimensional latent space acts as a bottleneck that restricts the flow of information through the network and, thus, when trained well forms an efficient representation of the input data. 

Auto-encoders have been used in astronomy in the past for applications such as denoising (training on pairs of noisy, near-noiseless, or augmented data; e.g., \citealt{2017A&A...603A..60F, 2018MNRAS.476.5365S, 2022MNRAS.509..990G}) , dimensionality reduction (e.g., \citealt{2020AJ....160...45P}), and representation learning (e.g., \citet{2020MNRAS.494.3750C} for identifying strong lenses and \citet{2021ApJ...913...12D} for chemical tagging). There have, to our knowledge, not been any astronomical applications of the type of encoder-decoder network with cross-domain data that we discuss in \secname\,\ref{subsec:endecode} below. 

Variational auto-encoders are a type of auto-encoder that incorporates variational inference into the model \citep{1312.6114}. Variational inference is a maximum-likelihood estimation (MLE) method for situations in which the probability density is very complex, which in our case is the distribution of the latent variables that generate the output data. The addition of variational inference in practice acts as a regularization that forces the latent space distribution to follow a given statistical distribution. This is often a Gaussian distribution, as any distribution can be described as a set of normally-distributed variables mapped via a complex function like a neural network. Moreover, forcing the latent space distribution to follow a known statistical distribution also makes generating new samples easier, because we can draw random samples from the latent space distribution and decode them to construction outputs. A detailed review of variational auto-encoders can be found in \citet{2016arXiv160605908D}.

Overall, a variational auto-encoder has the following major components: 

The \emph{encoder:} The encoder is a discriminative model that takes input data and compresses it to a latent space of much lower dimensionality. In variational auto-encoders, the encoder predicts the means and variances (in practice the encoder predicts logarithmic variance instead for numerical stability) of each latent-space parameter  and then samples from a normal distribution with these parameters to obtain the final efficient representation. 

The \emph{latent Space:} The latent space is the layer that is populated by the latent variables. The latent space is fully unconstrained prior to training and the entire latent-variable representation is learnt during training. However, we do need to set the dimension of the latent space, analogous to setting the number of principal components to use in PCA.

The \emph{decoder:} The decoder is a generative model that takes the latent variables and generates the (generally much higher dimensional) output. This is done by starting from the low-dimensional representation and building it out to the high-dimensional output through a sequence of steps (layers). In many ways, the decoder does the opposite of the encoder.

Variational auto-encoders are trained by minimizing an objective function that is composed of the sum of two loss terms. The first of these is the reconstruction loss, i.e., a measure of how well the model predicts the output. The second is a regularization term that shapes the distribution in the latent space, i.e., forcing the latent space to follow a certain distribution. We have adapted the mean squared error (MSE) as the reconstruction loss and the Kullback–Leibler (KL) divergence as the latent-space regularization loss to force the latent space to follow a Gaussian distribution. The MSE reconstruction loss $J_\mathrm{MSE}$ is given by
\begin{equation} \label{eq:mse}
J_\mathrm{MSE}(\vec{y}, \hat{\vec{y}}) = \frac{1}{N} \sum^N_{i=1} \vec{w}_i (\hat{\vec{y}_i}-\hat{\vec{y}_i})^2\,,
\end{equation}
where $\hat{\vec{y}_i}$ is the predicted output, $\hat{\vec{y}_i}$ is the true output, and $\vec{w}_i$ is a weight for each pixel $i$ in the output. The weight $\vec{w}$ will be an array of ones if no pixel-level weighting applied. The KL-divergence regularization loss $J_\mathrm{KL}$ is
\begin{equation} \label{eq:kl}
\begin{split}
J_\mathrm{KL}(\mu, \log\sigma^2) = \frac{1}{2}\left[-\sum_i\left(\log\sigma_i^2 + 1\right) + \sum_i\exp{(\log\sigma_i^2)} + \sum_i\mu^2_i\right]\,,
\end{split}
\end{equation}

where $(\mu_i,\log\sigma_i^2)$ are the mean and logarithmic variance in each latent-space dimension for data point $i$ from the encoder.

\begin{figure*}
\centering
\includegraphics[width=0.95\textwidth]{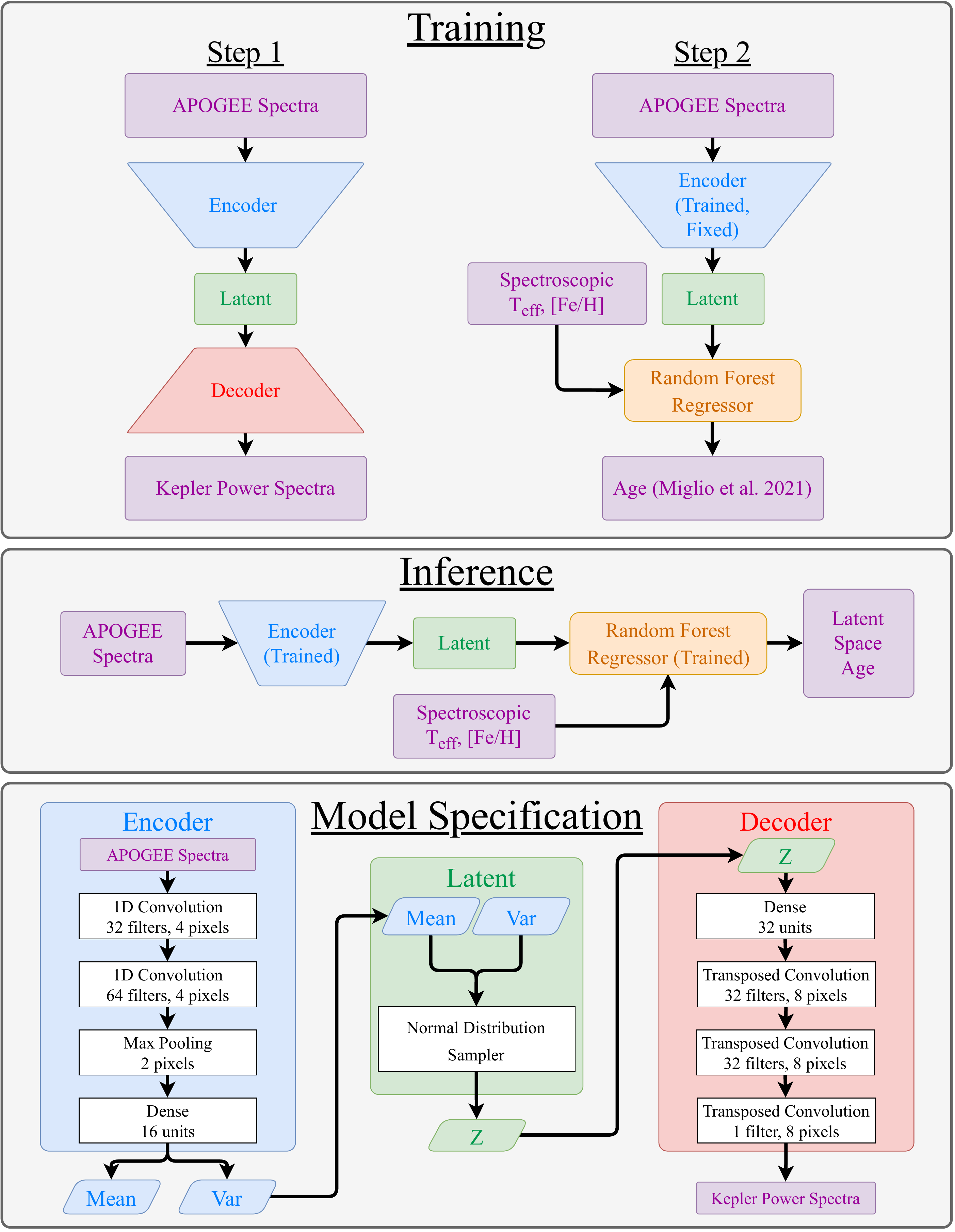}
\caption{Schematics of the methods and models that we use to obtain spectroscopic ages from a latent-space representation of APOGEE spectra. The top panels show the training and inference phase. In step 1 of the training phase, the encoder and decoder act as a single model that takes APOGEE spectra as inputs with the objective to reconstruct \kepler\ PSDs as outputs. In step 2, we train a random forest regressor to determine age from the latent space of the encoder-decoder along with \teff\ and \xh{Fe}. During inference, we discard the decoder and use the encoder to compress APOGEE spectra to the latent space, and then on through the trained random forest regressor to get the age prediction. The bottom panel specifies the detailed architectures of the encoder and the decoder. Both the encoder and the decoder are essentially convolutional neural networks.}
\label{fig:model_diag}
\end{figure*}

\subsection{Encoder-Decoder networks}\label{subsec:endecode}

A encoder-decoder network is very similar to an auto-encoder in terms of architecture and the same applies to variational encoder-decoders with respect to variational auto-encoders. They both consists of three major components, an encoder, a decoder and a latent space between the encoder and the decoder as we have discussed in the previous sub-section. The major difference between the two is in the input and output data used by the model. An auto-encoder uses data in the same domain (e.g., pairs of images of same objects) for the input and output node, while an encoder-decoder network employs data in different domains (e.g., image input and text output). Encoder-decoders are commonly used in neural machine translation (e.g., \citealt{2014arXiv1409.1259C, 2017arXiv170603762V}), because different languages (hence different domains) are just different ways to express the same abstract ideas.

In our application of an encoder-decoder in this paper, we have two domains of data, which are APOGEE high-resolution spectra as inputs on the one hand and power spectral density (PSD) derived from \kepler\ light-curves as outputs on the other hand. The goal of using an encoder-decoder is to extract the information about the \kepler\ PSDs that is contained in the APOGEE spectra, that is, the ``asteroseismic'' information that is present in the spectroscopic data. We do not necessarily extract the true asteroseismic information from APOGEE spectra, but simply any spectral information that is correlated with the asteroseismic information. However, because \kepler\ PSDs do not contain abundance information, we emphatically do not extract abundance information at the same time (as we will explicitly demonstrate below). To be able to successfully construct \kepler\ PSDs, we expect that the latent space must contain information on the asteroseismic parameters $\numax$ and $\Dnu$ from which the mass can be derived using scaling relations \citet{1995A&A...293...87K, 1991ApJ...368..599B} as
\begin{equation} \label{eq:scaling}
\frac{M}{M_\odot} \simeq \Big(\frac{\numax}{\nu_\mathrm{max,\odot}}\Big)^3 \Big(\frac{\Dnu}{\Dnu_{\odot}}\Big)^{-4}\Big(\frac{T_{\rm eff}}{T_{\rm eff,\odot}}\Big)^{3/2}\,,
\end{equation}
where $\nu_\mathrm{max,\odot}$ and $\Dnu_{\odot}$ are the asteroseismic parameters of the Sun. Thus, because spectra also contain information about $T_{\rm eff}$, we expect the latent space to contain information on the stellar mass, from which the age can be derived. Note that we do not use the scaling relations anywhere in our methodology, but we simply use them here to argue that the encoder-decoder should be able to extract mass information from the high-resolution spectra if it can successfully reconstruct \kepler\ PSDs from stellar spectra.

\subsection{Rationale of using an encoder-decoder network} \label{subsec:rationale}

In this work, we use an encoder-decoder to map high-resolution infrared spectra from APOGEE to PSDs derived from \kepler\ light curves. This has the following advantages:
\begin{enumerate}
  \item \textit{Ability to train on all APOGEE/\kepler\ pairs:} When training the encoder-decoder, we do not need labels (i.e., mass or age). Thus, all that we require are stars that have both APOGEE spectra and \kepler\ light curves and we can use  \textit{all} overlapping observations between the APOGEE spectra and \kepler\ (or, in the future, \tess) light curves. There are many more APOGEE-\kepler\ pairs than there are reliable asteroseismic measurements of mass and age for those pairs, so this allows for a large expansion of the available training data.
  \item \textit{Information extraction:} Current data-driven spectroscopic ages likely rely on proxies like the [$\mathrm{C/N}$] ratio \citep{2016ApJ...823..114N, 2019MNRAS.489..176M} and they may even use information such as that contained in $\xfe{\alpha}$. However, the [$\mathrm{C/N}$] ratio is not only affected by mass-dependent mixing processes, but instead at least in part depends on galactic chemical evolution and this adversely affects age predictions. However, we have millions of spectra from surveys such as APOGEE and GALAH, so being able to determine ages without relying on abundance information beside \xh{Fe} is important. By using an encoder-decoder, we force the method to extract only information necessary to predict the \kepler\ PSD without extracting unnecessary abundance information. Information such as [$\mathrm{C/N}$] is not directly available in PSD, but its component that is due to stellar mixing may still emerge in the latent space through its correlation with mass.
  \item \textit{Application to large spectroscopic datasets:} Once trained, we can discard the decoder and apply the encoder plus the latent space to all available spectroscopic data, which cover the entire sky.
  \item \textit{Simplicity:} The dimensionality of the latent space is orders of magnitude smaller than that of the spectra and the PSD. Thus, we can train simpler regression models to predict the age from the latent space. Simple models can be trained using only a handful of very precise ages.
\end{enumerate}

\begin{figure*}
\centering
\includegraphics[width=0.95\textwidth]{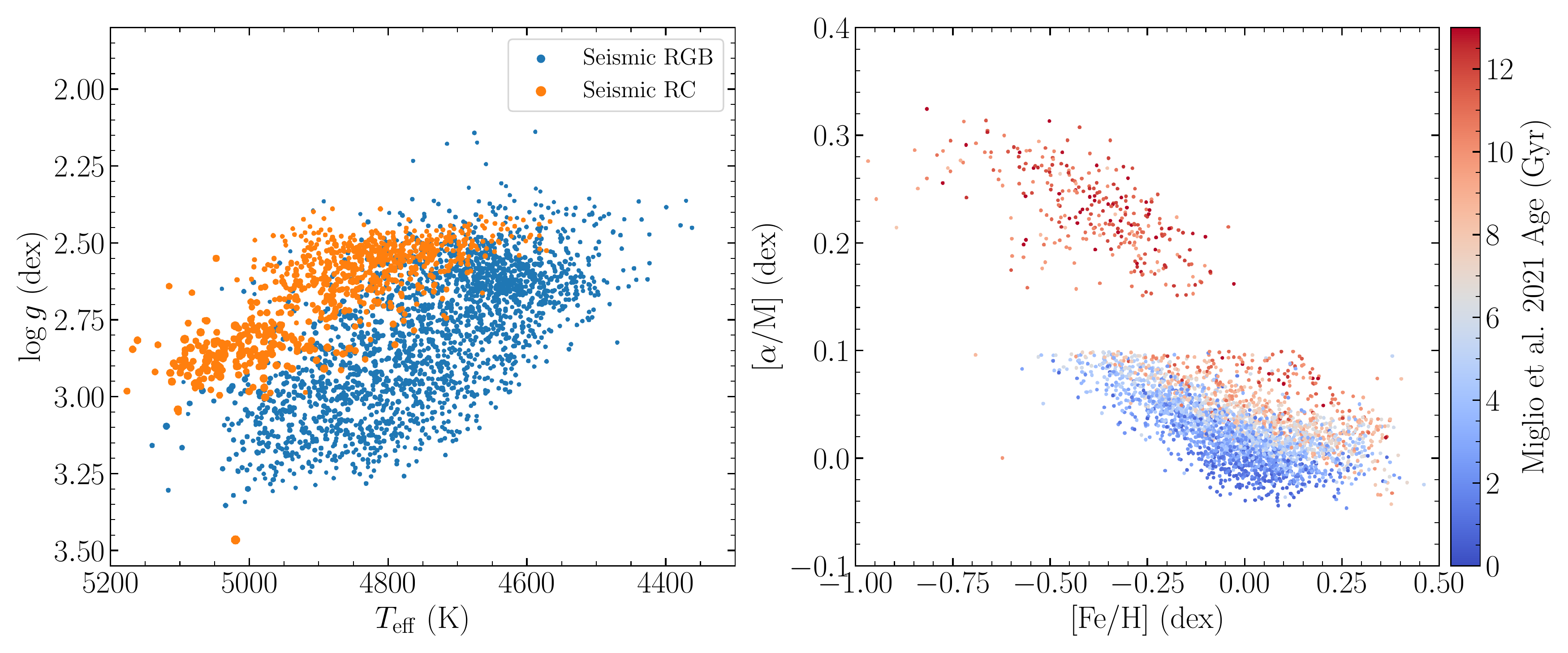}
\caption{Age measurements from \citet{2021A&A...645A..85M} which are used for training in this paper. The left panel shows the distribution of stars in \teff-\logg\ space colored by their seismic classification of red-giant branch (RGB) or red-clump (RC) stars with the marker size corresponding to their mass. The \logg\ has a limited range with the majority of the stars lying within $2.5 \lesssim \logg  \lesssim 3.3$ dex. The right panel show the distribution of stars in \xh{Fe}-\xfe{\alpha} space colored by age. There is an artifact at \xfe{\alpha} at 0.1 dex to separate $\alpha$-rich and $\alpha$-poor population, as \citet{2021A&A...645A..85M} separate their analysis into $\alpha$-rich and $\alpha$-poor stellar samples.}
\label{fig:miglio_data}
\end{figure*}

\section{Models} \label{sec:model}

In this section, we provide details on the encoder-decoder network that we use to map APOGEE spectra to \kepler\ PSDs and on the regressor that we employ to map the latent space to age. A schematic overview of our methodology is shown in \figurename\,\ref{fig:model_diag}. The encoder-decoder part is implemented as \texttt{ApokascEncoderDecoder()} using \texttt{tensorflow} \citep{tensorflow2015-whitepaper} in the \texttt{astroNN} package \citep{2019MNRAS.483.3255L} \footnote{\url{https://github.com/henrysky/astroNN}}.

\subsection{Encoder-decoder model} \label{subsec:endecoder}

The encoder in our model consists of a convolutional neural network with two convolutional layers, a max-pooling layer, and a dense layer that outputs the mean and variance for each dimension of the latent space starting with APOGEE stellar spectra. We use the \texttt{ReLU} activation function throughout the encoder except in the final layer where we employ the \texttt{tanh} activation function to prevent predicting extreme values in the latent space and improve training stability. The latent space is simply a normal distribution sampler that samples latent variables from the output of the encoder. The decoder is another convolutional neural network, but with transposed convolutional layers to reconstruct the \kepler\ PSD as opposed to the regular convolutional layers used in the encoder. We again use the \texttt{ReLU} activation function throughout the decoder, except in the final layer where no activation is applied to the outputs in order not to limit the range of the reconstruction output.

The conventional convolutional layer \citep{LeCun6795724} used in convolutional neural networks excels in pattern recognition. It works by learning convolutional filters that recognize certain patterns, hence the output values are calculated as the dot product between filter and the input to the layer by moving the filter kernel across the input, which is the usual convolution operation. For the transposed convolutional layer (also known as deconvolution), the input value in the layer determines the filter values that will be written to the output. In other words, the input determines the weight of the filters as opposed to learning the weights as in the usual convolutional layers.

Hyper-parameters such as the number of neurons, the dimension of the latent space, and the optimizer's learning rate are optimized by hyper-parameter search. For parameters like the latent space dimension, we have an educated guess of what value we want to use. We know at least three parameters are important to the reconstruction of the PSD: \numax, \Dnu, as well as the evolutionary state. From this minimum number of dimensions, the latent space dimension is increased until there is no increase in the performance for the validation set. This gives a five-dimensional latent space.

\subsection{Probabilistic random forest model} \label{subsec:rf}

To map the latent space to stellar age, we employ a version of the random forest implementation from \texttt{scikit-learn} \citep{2012arXiv1201.0490P}. In \texttt{scikit-learn}, random forests are built with individual trees in the forests in which each tree is built from bootstrapped samples from the training set; the prediction is simply the mean of the value returned by each tree in the forest. We have modified the implementation such that during training, we sample the input and output data with their corresponding uncertainty on top of the bootstrap sampling where the input data uncertainties are taken directly from the encoder (i.e. variance predicted by the encoder) while the output data uncertainties are given in training set. Each star is then weighted by the age uncertainty of the training set by $\sqrt{1/\sigma_{\mathrm{age}}^2}$. At testing time, the predicted age is the mean of all of the trees and we interpret the standard deviation of the trees as the age uncertainty.

\section{Datasets and Data Reduction} \label{sec:data}

We use spectroscopic data from APOGEE, light curves and derived PSDs from \kepler, and ages for stars in the APOKASC catalog derived by \citet{2021A&A...645A..85M}. We discuss these different data sets in this section and describe the relevant data reduction processes for each dataset.

\subsection{APOGEE}\label{subsec:apogee}

We use high-resolution spectra from the APO Galactic Evolution Experiment (APOGEE; \citealt{2017AJ....154...94M, 2017AJ....154...28B}), specifically from its seventeenth data release (DR17; \citealt{2022ApJS..259...35A}). APOGEE DR17 is a high signal-to-noise ($>100$ per pixel typically), high resolution ($R\sim22,000$) panoptic spectroscopic survey in the near infrared H-band wavelength region of $1.5-1.7\um$. APOGEE spectra are the input of our neural network model and we employ the same continuum normalization procedure as in \citet{2019MNRAS.483.3255L}. In addition to APOGEE spectra, we use stellar parameters and elemental abundances derived by the APOGEE Stellar Parameter and Chemical Abundances Pipeline (ASPCAP; \citet{2016AJ....151..144G}). These are not used by the encoder-decoder part of our model, but we add some of them to the input data given to the random-forest regressor when mapping the latent space to age. Specifically, we add the effective temperature \teff\ and the metallicity \xh{Fe}, because stellar models require these to determine stellar ages from asteroseismic observations. We use the ASPCAP parameters rather than external data-driven stellar parameters and elemental abundances such as those from \citet{2019MNRAS.483.3255L} or \citet{2019ApJ...879...69T} that are proven to be robust to lower signal-to-noise ratio spectra, to be consistent with the stellar parameters used to derive the training ages.

\begin{figure*}
\centering
\includegraphics[width=0.95\textwidth]{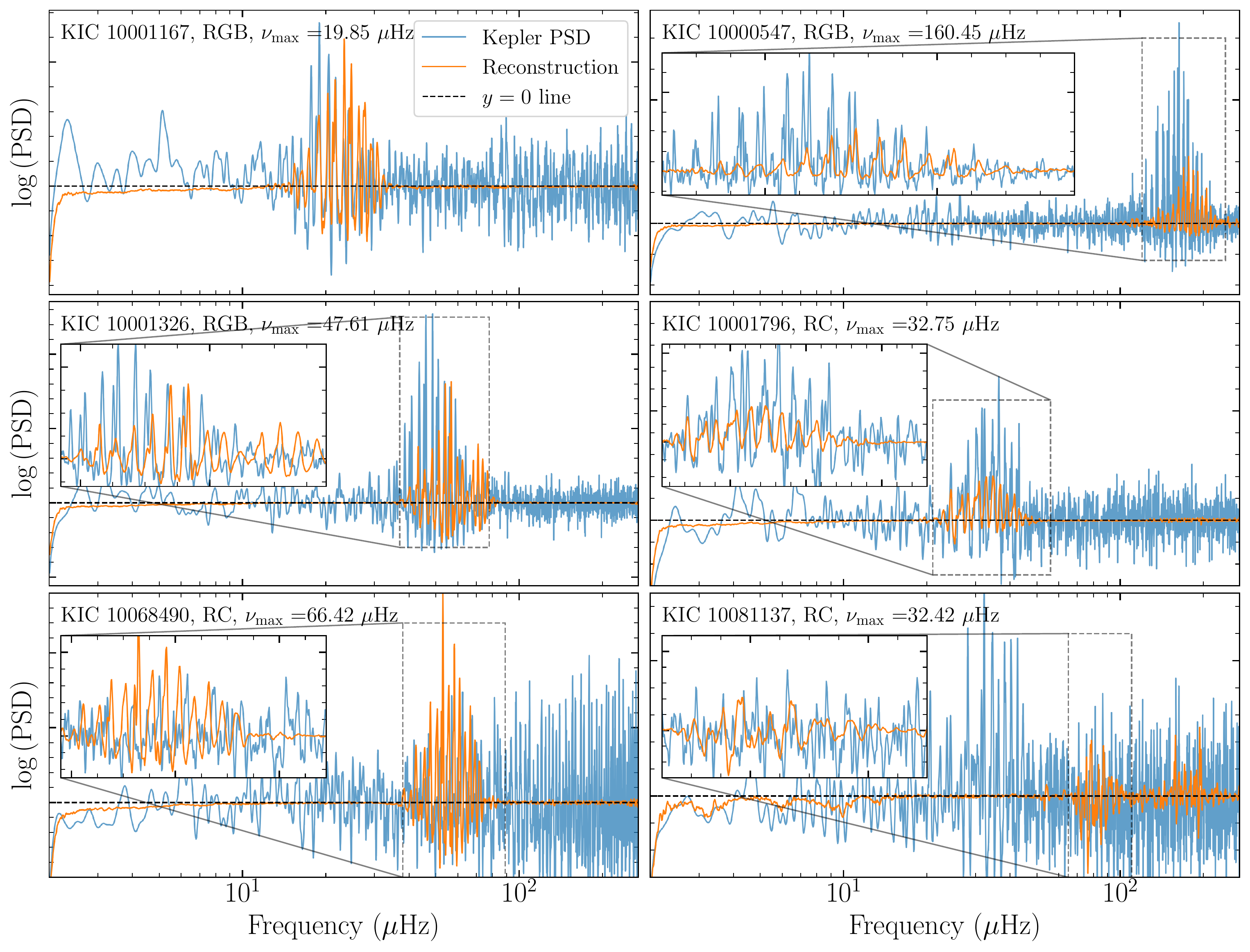}
\caption{Reconstruction of the power spectral density (PSD) from APOGEE spectra with our encoder-decoder model (orange line) compared to the PSD generated directly from the \kepler\ light curve (blue line) for stars not included in training set, such that the model has never seen those spectrum-PSD pairs before (except for the bottom right panel example, which is from the training set). Our encoder-decoder model successfully reconstructs the area of interest where the p-mode envelope is. The top-left panel demonstrates that even in the case of low \numax, where the amplitude of the p-mode envelope relative to the background is low, the model still successfully reconstructs the envelope at approximately the correct frequency location. The bottom two panels show two RCs with peculiar PSD reconstructions. The bottom left one shows one RC star that is not included in the training set where the p-mode envelope is reconstructed at the correct \numax, but the overall shape does not resemble the ground truth PSD. The bottom right panel show an RC star \textit{included} in the training set where the model fails to reconstruct the p-mode envelope at the expected \numax. In general, the neural network correctly identifies and reconstructs the p-mode envelope and it does not reconstruct the noise, which is expected as noise is unpredictable from the APOGEE spectra.}
\label{fig:recon}
\end{figure*}

\subsection{Kepler}\label{subsec:kepler}

We use light curves for giant stars obtained by the \kepler\ telescope \citep{2010Sci...327..977B} in the original Kepler field near the Cygnus and Lyra constellations, to computer the power spectral densities (PSD) that are the output node of our encoder-decoder model. To download, manage, and manipulate \kepler\ data, we make use of the \texttt{lightkurve} Python package \citep{2018ascl.soft12013L}. We compute the PSD from the observed light curve using the Lomb-Scargle periodogram \citet{1976Ap&SS..39..447L, 1982ApJ...263..835S} as implemented in \textit{astropy} \citep{2022ApJ...935..167A}. We adopt a minimum frequency of $2\,\uHz$ and maximum frequency of $270\,\uHz$ (roughly corresponding to half of the inverse of the \kepler\ 30 minutes cadence) with a frequency spacing of $0.009\,\uHz$ (roughly corresponding to the inverse of \kepler's 3.5 years baseline for our sample). These values are similar to those used in the Kepler Light Curves Optimized For Asteroseismology (KEPSEISMIC; \citealt{2011MNRAS.414L...6G}) work. We divide the PSD by a low-pass background filter largely corresponding to the noise background coming from stellar activity, granulation, and photon noise; the low-pass filter consist of a moving median filter with a width of $0.01$ in logarithmic $\uHz$ frequency space.

In this work, we employ the PSD instead of the auto-correlation function (ACF) used in works like \citet{2013MNRAS.432.1203M} and \citet{2018MNRAS.474.2094A} for asteroseismic data analysis as well and by \citet{2018ApJ...866...15N} for determining data-driven stellar parameters. The reason why the latter work prefers to use the ACF over the PSD is the smooth gradient with respect to parameters such as age of each pixel in the ACF, which is essential for models that require smooth gradients. Although our decoder is a generative model for the PSD, we do not use true stellar labels to generate the PSD but instead rely on  latent variables. Moreover, \citet{2020arXiv200509682B} demonstrated that using the PSD rather than the ACF in a discriminative model improves performance. In this paper, using the PSD improves interpretability of our model compared to the ACF as we can directly check whether the decoder reconstructs the expected asteroseismic modes in the PSD which will be shown and discussed in \secname\,\ref{subsec:psd_recon}.

The PSDs constructed using the method above have a large number of pixels and to reconstruct them with a neural network would require a large number of parameters that would likely be overfit due to the limited number of APOGEE-\kepler\ pairs available for training the encoder-decoder. Moreover, it is unlikely that the stellar spectra contain such extremely precise asteroseismic information that they can predict the PSD at  $0.009\,\uHz$ resolution. The resolution of interest for our application is proportional to the large frequency separation $\Dnu$ --- the average frequency spacing between modes of adjacent radial order of a given angular degree $l$. In particular, we want to resolve mass-dependent variations in $\Dnu$ at a given $\numax$. These are a few percent of the $\Dnu$ expected for a typical star at a given $\numax$ (see Equation \ref{eq:excessDnu} below). Thus, we rebin the PSD using a logarithmic spacing chosen such that $\Dnu$ is resolved by about 50 pixels at any given $\numax$ (i.e., on average there are 50 pixels between two oscillation modes of consecutive overtones of the same angular degree for stars with any \numax). This gives $2,092$ pixels in each power spectrum.

The frequency values $\mathord{\mathit{f}}_i$ of our final PSDs  (that is, the frequency $\mathord{\mathit{f}}_i$ in \uHz\ represented at every pixel $i$) are approximately given by the expression
\begin{equation} \label{eq:approx_freqsol}
\begin{split}
\mathord{\mathit{f}}_i = 77.35 \Big(\frac{i}{2092}\Big)^4 + & 102.55 \Big(\frac{i}{2092}\Big)^3 + 68.96 \Big(\frac{i}{2092}\Big)^2 \\ & + 18.47 \Big(\frac{i}{2092}\Big) + 2.00\,.
\end{split}
\end{equation}
The exact solution that we use is computed using the following recurrence relation
\begin{equation} \label{eq:exact_freqsol}
\mathord{\mathit{f}}_{i+1} = \frac{0.263\uHz\big(\mathord{\mathit{f}}_{i}^{0.772}\big)}{50} + \mathord{\mathit{f}}_{i}\,,
\end{equation}
where $\mathord{\mathit{f}}_0$ is $2\,\uHz$. The maximum deviation between the approximation and full solution is $0.008\,\uHz$ while median absolute deviation is $0.002\,\uHz$.

\begin{figure*}
\centering
\includegraphics[width=0.95\textwidth]{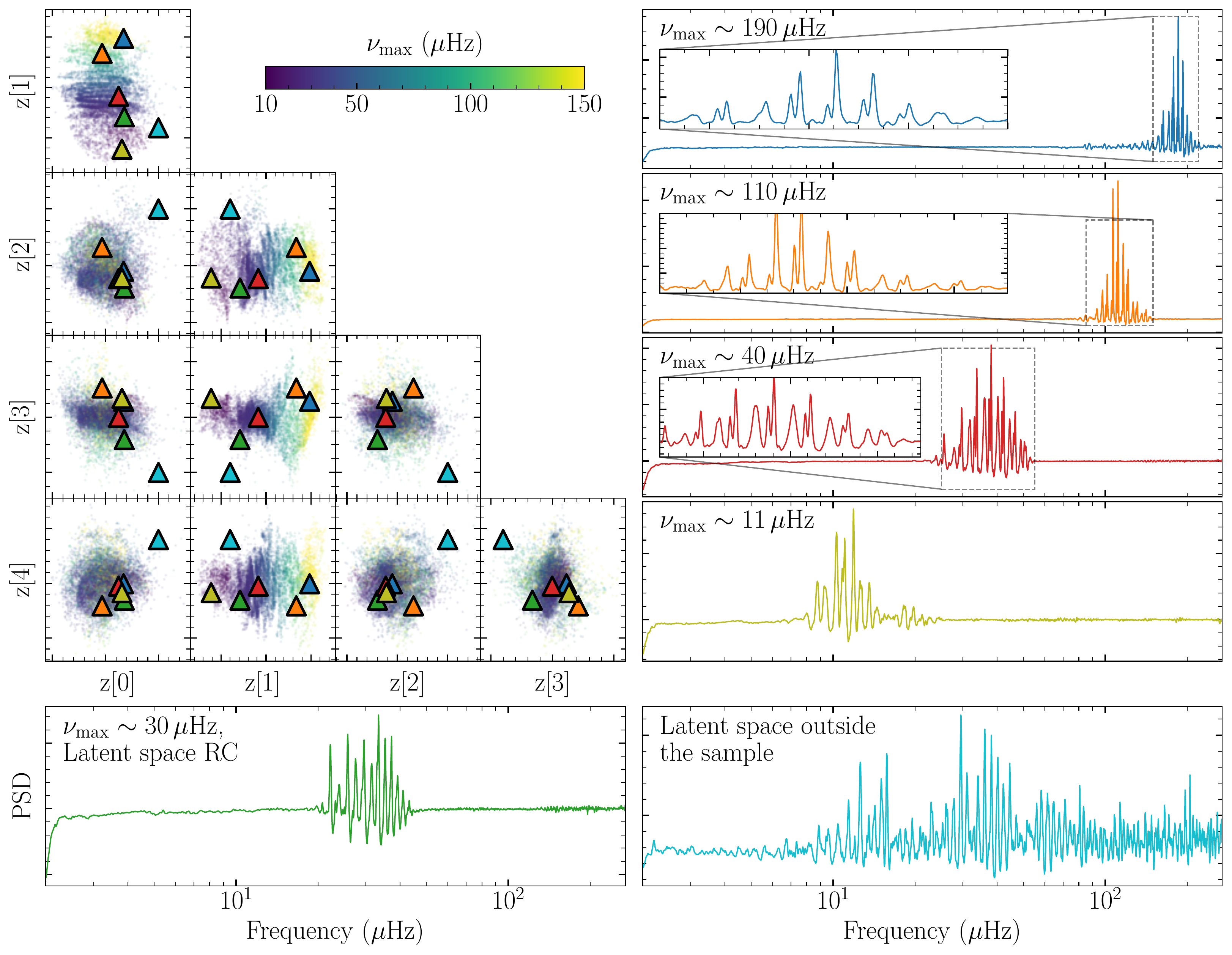}
\caption{Reconstruction of power spectral densities from a few locations in the latent space. The big panel (top-left) shows the latent space colored by \numax\ for the encoder-decoder training sample (i.e., the top-right panel of \figurename\,\ref{fig:latent_colored}). The reconstructed PSDs at the locations of the triangles are shown in the six smaller panels, where the colors of the lines correspond to the color of the triangle markers in the big panel. The four top right panels show reconstructions for PSDs with \numax\ around $190\,\uHz, 110\,\uHz, 40\,\uHz$ and $11\,\uHz$. The bottom left panel displays a PSD reconstruction from a region associated with the RC. The bottom right panel shows 
a reconstruction from a region not populated by any real-world APOGEE spectra; this PSD is much noisier than the others. All PSD panels have arbitrary units on the y-axis and logarithmic frequency the x-axis.}
\label{fig:latent_recon}
\end{figure*}

\subsection{APOKASC and Ages}\label{subsec:apokasc}

For asteroseismic data, we use data from the APOGEE-\kepler\ Asteroseismology Science Consortium (APOKASC; \citealt{2014ApJS..215...19P}) catalog, the \citet{2018ApJS..236...42Y} \kepler\ red-giant seismology catalog, and ages from \citet{2021A&A...645A..85M}. The APOKASC catalog consists stars observed by both the APOGEE survey and the \kepler\ telescope. We have adopted the second data release of the catalog (APOKASC-2; \citealt{2018ApJS..239...32P}) based on APOGEE DR14 \citep{2018ApJS..235...42A}. The APOKASC-2 catalog consist of $6,676$ giant stars, $85\%$ of which have \kepler\ light curves with time baseline of $>3.5$ years. APOKASC first estimates global seismic parameters \numax\ and \Dnu\ from power spectra generated from the \kepler\ light curves using multiple pipelines that suffer from different systematics and apply different constraints on the model parameters. To expand the sample size of APOGEE-\kepler\ pairs, we cross-match the \citet{2018ApJS..236...42Y} catalog with APOGEE DR17 to obtain $10,526$ stars. We use the union between the APOKASC-2 and \citet{2018ApJS..236...42Y} catalog to obtain $10,672$ stars with solar-like oscillations.

The global parameters \numax\ and \Dnu\ are highly correlated and we adopt the following relation that describes this correction when relevant
\begin{equation} \label{eq:numax_dnu}
\Dnu = 0.263 \uHz \times (\numax / \uHz)^{0.772}\,.
\end{equation}
which is a relation that is empirically determined using results from the \texttt{SYD} asteroseismic pipeline \citep{2009CoAst.160...74H, 2010ApJ...723.1607H}. We parameterize deviations from this relation for individual stars using what we call ``excess \Dnu", defined as
\begin{equation} \label{eq:excessDnu}
\mathrm{Excess}\ \Dnu\ = \Dnu \big/ \big( 0.263\uHz\, [\nu_\mathrm{max}/\uHz]^{0.772} \big)\,,
\end{equation}
where \numax\ and \Dnu\ are the measured values for a given star. Excess \Dnu\ can be shown to be sensitive to stellar mass using the asteroseismic scaling relations (a review on asteroseismology and seismic scaling relations can be found in \citealt{2019LRSP...16....4G}).

To train the latent-space-to-age part of our model, we adopt training ages from \citet{2021A&A...645A..85M}. Specifically, we use a version of the method from this paper that is updated to use the DR17 ASPCAP stellar parameters and abundances as opposed to those from DR16 in the original work. The \citet{2021A&A...645A..85M} catalog consist of $3,078$ evolved stars with measured ages. The distribution in the $\logg$-$\teff$ plane and in the $\alpham$--\xh{Fe} plane of the sample is given in \figurename\,\ref{fig:miglio_data}. The \citet{2021A&A...645A..85M} method uses a similar approach as the APOKASC catalog to determine masses, radii, and ages using the observed light curves, spectroscopic parameters from APOGEE, and stellar models, but it has small implementation differences. \citet{2021A&A...645A..85M} uses \texttt{PARAM} \citep{2006A&A...458..609D, 2017MNRAS.467.1433R} to infer masses and ages while APOKASC-2 uses \texttt{BeSPP} \citep{2017ApJS..233...23S}. Another difference is that \citet{2021A&A...645A..85M} determines the global seismic parameters using peak-bagging \citep{2016MNRAS.456.2183D}, where individual radial-mode ($\ell=0$) frequencies are fit for a subset of stars to get \numax\ and \Dnu, allowing a better measurement of those parameters. This in turn leads to improved stellar ages. The ages from \citet{2021A&A...645A..85M} that we use have a standard deviation of $\sim 10\%$ compared to ages in the APOKASC-2 catalog. In general, the \citet{2021A&A...645A..85M} age uncertainty is $\sim 5\%$ larger for RGB but $\sim 15\%$ less for RC stars when compared to APOKASC-2. 

\section{Training and Testing} \label{sec:traintest}

Before training, we further standardize the APOGEE spectra by subtracting the pixel-level mean and dividing by the pixel-level standard deviation after the data reduction steps discussed in \secname\,\ref{subsec:apogee}. We restrict output PSDs to have $4 < \numax < 250\,\uHz$ to ensure that the whole p-mode power envelope can be seen as a whole in the PSD with $\sigma_{\numax}$ less than $10\%$. We restrict the sample of APOGEE-\kepler\ pairs to those with PSDs with evolutionary state determinations from APOKASC-2 or \citet{2018ApJS..236...42Y}, as this provides a good indication of the quality of the PSD (some PSDs without evolutionary state determinations have \numax\ that appear far off visually). We predict the logarithmic amplitude of the PSD. The first step of the training process is to train the encoder-decoder (as shown in \figurename\,\ref{fig:model_diag}) as a whole with $9,869$ pairs of APOGEE spectra and \kepler\ PSDs randomly selected from all APOKASC pairs. We optimize the model using the \texttt{ADAM} optimizer \citep{2014arXiv1412.6980K}. To accelerate training as well as to make the training process more stable, we weight pixels near the observed \numax\ higher when calculating the objective function in \eqnname\,\ref{eq:mse}. We generate a Gaussian normalized to a height of one for each PSD centered at \numax\ with a width of twice the \Dnu\ expectation from \eqnname\,\ref{eq:numax_dnu} and we add this Gaussian to the existing sample weights for each pixel which were arrays of ones. It is not necessary to know \numax\ in advance and apply this pixel-level weighting to obtain convergence to a good model, but the addition of this weighting makes the model converge much faster. The introduction of this weighting scheme does not prevent our model from learning features outside of the p-mode power envelope as shown in the bottom right panel of \figurename\,\ref{fig:recon}.

After training the encoder-decoder, APOGEE spectra in the APOKASC sample are run through the trained encoder (but not the decoder) to get their latent representations. We then train random forest models to go from these latent representations to other labels, primarily age, but we also predict other quantities from the latent space to aid in understanding the latent space; this training step is shown in the middle panel of \figurename\,\ref{fig:model_diag}. When predicting chemical abundances and global seismic parameters from the latent space, $2,000$ stars are randomly selected to be the training set. When predicting stellar ages, about $1,200$ stars are randomly selected among the $2,000$ stars with ages to be the training set. Our fiducial model predicts age from the latent space augmented with \teff\ and \xh{Fe}; we refer to these as ``latent space ages''. In practice, we predict the logarithm of the age and we do not apply any bounds on this logarithm.

\section{Results} \label{sec:result}

The ultimate goal of our method is to determine ages from the high-resolution APOGEE data using the encoder, the latent space, and the random-forest regressor from the latent space to age. Because our methodology consists of multiple important components that build on each other to provide the final ages, we discuss the results from the components in order in this section. We commence with a discussion of the performance of the encoder-decoder in reconstructing the PSD from spectra in \secname\,\ref{subsec:psd_recon}, then we take a detailed look at the latent space in \secname\,\ref{subsec:latent}, as well as describe how well we can predict different stellar parameters, including age, from the latent space representation of the spectra in \secname\,\ref{subsec:realresult}. Finally, we discuss the application of our model to the whole APOGEE DR17 data set to get ages for a large sample of stars.

\begin{figure*}
\centering
\includegraphics[width=0.95\textwidth]{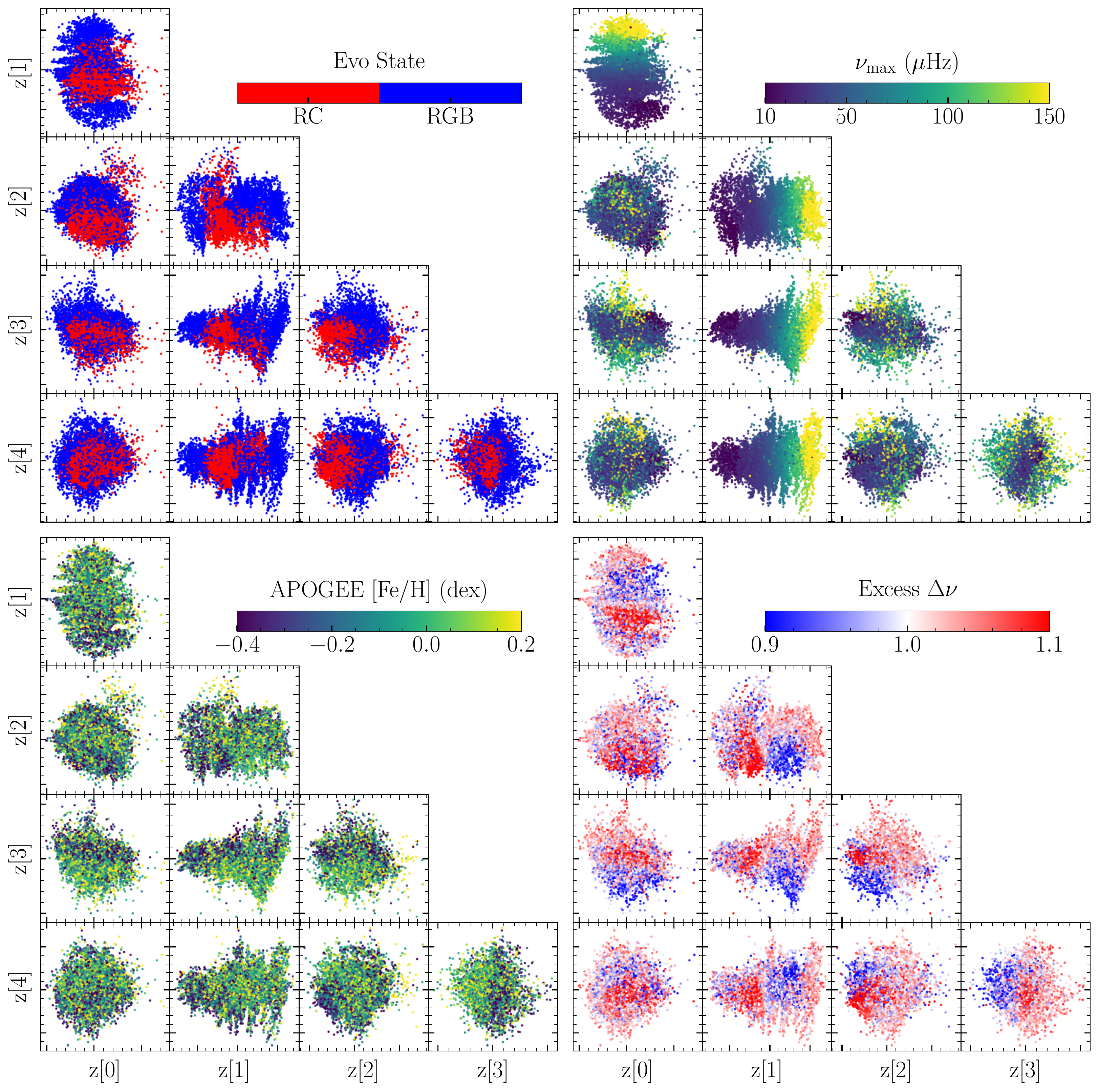}
\caption{This figure shows the latent space prediction of our encoder-decoder model running through all the APOGEE spectra in the encoder-decoder training set and colored by different labels (taken from external catalogs, not predicted by our neural networks). The top left panel shows the latent space colored by evolutionary state, with blue markers for RGB stars and red markers for RC stars. RC stars clearly cluster together in most of the latent space dimension. The top right panel show the same latent space but colored by \numax. Smooth trends in \numax\ are clearly seen in most of the dimensions, which is expected as the encoder-decoder is able to reconstruct the PSD so well as shown in \figurename\,\ref{fig:recon}. The bottom right panel displays the latent space colored by excess \Dnu, demonstrating that the model indeed learns the tiny shift in those oscillation peaks related to stellar mass instead of just generating visually good-looking PSDs. The bottom left panel shows the latent space colored by \xh{Fe}. There is no clear \xh{Fe} trend in any dimension, showing that the model did not learn \xh{Fe}. This figure is discussed in detail in \secname\,\ref{subsec:latent}.}
\label{fig:latent_colored}
\end{figure*}

\begin{figure*}
\centering
\includegraphics[width=0.95\textwidth]{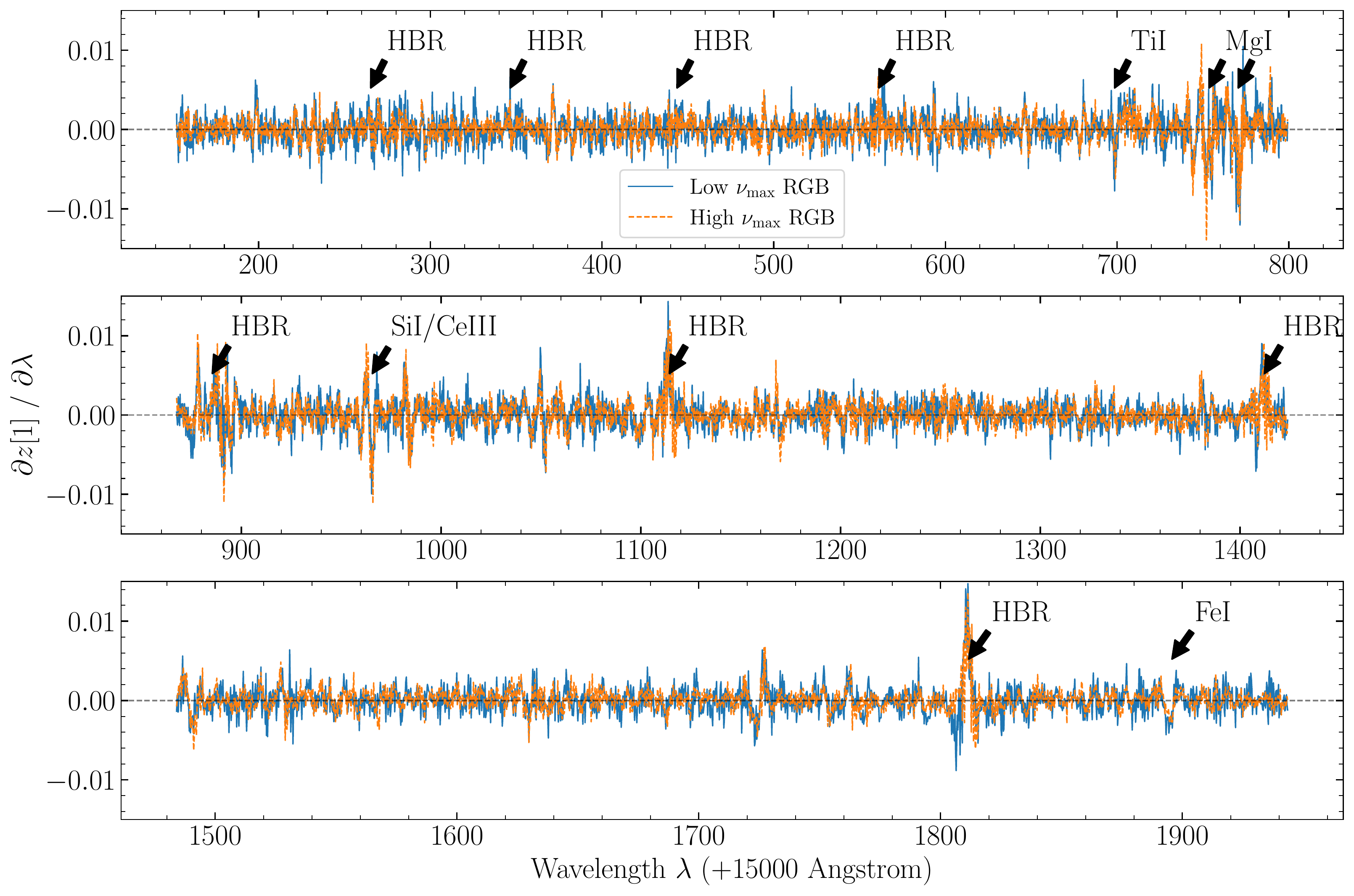}
\caption{Jacobian of the second dimension z[1] of the latent space for each pixel in the spectra (i.e., how changes in each pixel in the APOGEE spectrum changes the value of z[3]) for RGB stars with low \numax\ (blue line; $15\,\uHz \lesssim \numax \lesssim 20\,\uHz$) and high \numax\ (orange dashed line; $180\,\uHz \lesssim \numax \lesssim 220\,\uHz$) using the encoder part only in the encoder-decoder model. There are a few regions, especially around the hydrogen lines (the Brackett series) in the green and red regions of the APOGEE spectra that are known to be sensitive to \logg, with large sensitivity for the low \numax\ RGB stars but little sensitivity for the high \numax\ RGB stars in this particular latent space dimension. Overall, the information about the latent space appears to be spread over the entire spectrum.}
\label{fig:latent_jacob}
\end{figure*}

\subsection{Power spectral density reconstruction}\label{subsec:psd_recon}

After training the encoder-decoder on the APOGEE/\kepler\ pairs, we can directly check the performance of the model by generating the reconstructed PSDs for spectra in a test set of pairs that was not used during the training. We expect that any information in the PSD that is predictable from (or correlates with) the APOGEE spectra to be present in the output. PSD information that is not contained in the APOGEE spectra cannot be obtained by the decoder-encoder, instead the model will predict the mean value of those PSD pixels in the sample to minimize the objective function (\eqnname\,\ref{eq:mse}). Information in the APOGEE spectra that is not necessary to reconstruct the PSD will, similarly, not be part of the latent space (this crucial last point is discussed in detail in \secname\,\ref{subsec:latent}). 

In \figurename\,\ref{fig:recon}, we compare the actual power spectra computed from \kepler\ light curves and the encoder-decoder reconstruction from the APOGEE spectrum for pairs in the set of APOGEE/\kepler\ pairs that are not included in training or validation set and, thus, the model has never seen those pairs before. The figure includes stars with large and small \numax\ and of different evolutionary states. We see that, overall, the encoder-decoder reconstructs the p-mode envelope remarkably well. The p-mode envelop is the area of interest for our purposes, because it contains most of the seismic information relevant to age determination. The encoder-decoder is also able to reconstruct the locations and heights of individual oscillation modes to high precision considering the information comes from spectroscopic observations that only contain information on the surface condition of stars. Thus, the encoder-decoder appears to have learnt the global seismic properties like \numax\ and \Dnu.

The encoder-decoder is able to reconstruct the p-mode envelope well regardless of the value of \numax. The top-left panel of \figurename\,\ref{fig:recon} demonstrates that even for a star with low \numax\ ($< 20\,\uHz$), where the amplitude of the p-mode envelope relative to the background is low, the model still successfully reconstructs the envelope with an overall peak at approximately the correct location and with individual oscillation modes at the approximately correct locations as well (note that the absolute amplitude of the p-modes are higher at low \numax, but lower relative to the background granulation and activity noise so PSDs for low \numax\ stars seem "noisier" ). In general, the encoder-decoder correctly identifies the location of the p-mode envelope and of the individual modes for a wide range of stars and reconstructs them correctly. The model does not reproduce the noise in the PSD (e.g., the photon noise instrumental systematics, and stochastic granulation), which is expected because the noise is unpredictable from the APOGEE spectra. Increasing the latent space dimension from the five dimensions used by the model does not improve the PSD reconstruction for the testing set. The reconstructions are likely limited by the spectroscopic information available in APOGEE spectra.

The use of a variational method in the encoder-decoder also allow us to generate new PSD samples directly from the latent space (i.e., without starting from an APOGEE spectrum). In \figurename\,\ref{fig:latent_recon}, we show PSDs generated from latent space locations not directly associated with APOGEE spectra, but within the parameter space of the training set,  and one location outside of the training set's parameter distribution. The PSD samples generated look reasonable across a wide range of \numax, except except the one generated in latent space region associated with the RC, which does not have the correct envelope shape, and the one (cyan) that is generated outside of the parameter space of the training set, which is very noisy.

\begin{figure*}
\centering
\includegraphics[width=0.95\textwidth]{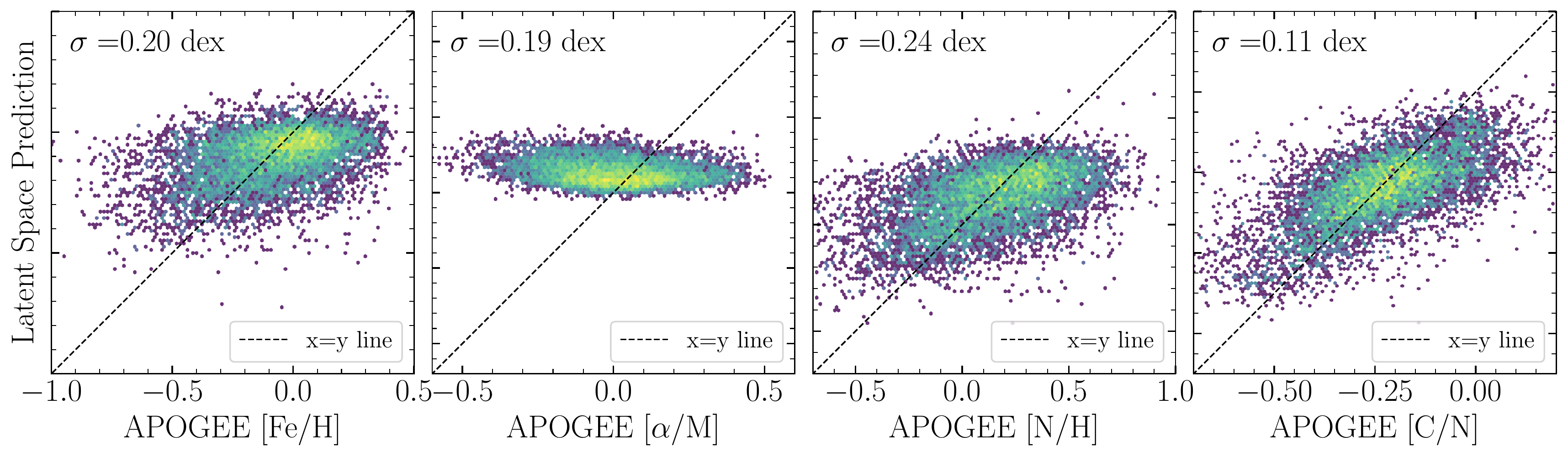}
\caption{Latent-space abundance predictions. This figure shows density plots on a logarithmic scale of APOGEE abundance prediction from the latent space compared to the ground truth for \xh{Fe}, \alpham, \xh{N}, \cn, from left to right respectively, with the overall scatter around the one-to-one line shown in the top-left of each panel. It is clear that the latent space contains no information on \alpham, a small amount of information on \xh{N} and \xh{Fe}, and some information on \cn.}
\label{fig:latent_aspcap}
\end{figure*}

\begin{figure*}
\centering
\includegraphics[width=0.95\textwidth]{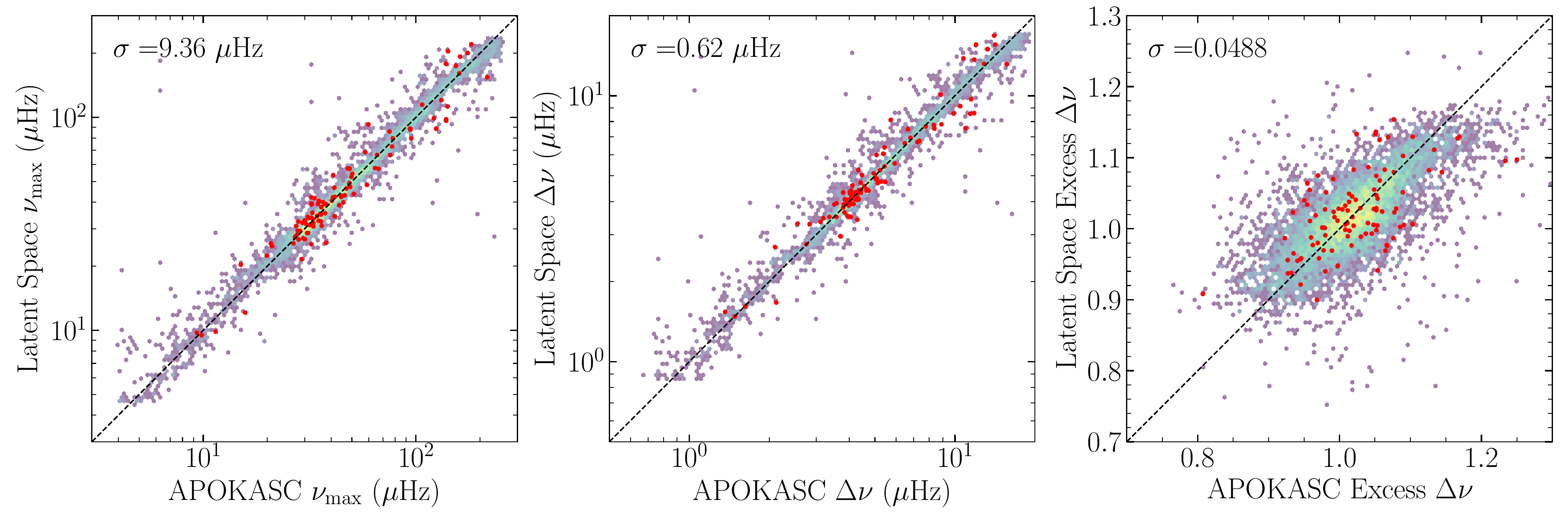}
\caption{Latent-space seismic parameter predictions. This figure shows density plots on a logarithmic scale of the global seismic parameters predicted using the latent space compared to the ground truth for \numax\ in the left panel, \Dnu\ in the middle panel, and excess \Dnu\ in the right panel. These stars are not contained in the training set for the latent space model, but some are contained in the encoder-decoder training set. The red points show additional stars not included in the training set for either the latent space model or the encoder-decoder model; these red points follow approximately the same distribution as the density plot. We are able to determine the important seismic parameters from the latent space to high precision.}
\label{fig:latent_seismic}
\end{figure*}

\subsection{The latent space} \label{subsec:latent}

The latent space of the trained encoder-decoder contains the information learnt by the encoder in its quest to reconstruct the PSD from the near-infrared APOGEE spectrum. It consist of the information that the model found to be crucial to predicting the PSD. Unlike checking the performance of the encoder-decoder as a whole by validating the quality of the PSD reconstruction using test data, the latent space is unsupervised in that we do not use any data to directly inform or restrict its structure. The only constraint placed on the latent space is the variational constraint that pushes it towards having a Gaussian distribution (see \secname\,\ref{subsec:autoencoder}). Thus, the best we can do is to study the latent space and make sure that it has properties that we expect and desire when going from spectra to PSD: (i) that it clearly encodes crucial seismic information such as \numax\ and \Dnu\ and (ii) that it does not contain extraneous spectral information such as elemental abundance ratios that are unnecessary for reconstructing the PSD but that have galactic-evolution correlations with age. 

We find that we can satisfactorily predict the PSD using a five-dimensional latent space. \figurename\,\ref{fig:latent_colored} displays the distribution of the data in the five-dimensional latent space values for all APOKASC stars whether they are in the training set or not (we can use all data, because we are simply trying to understand what the model has learned as opposed to directly validating the model, and the large number of stars in the full sample makes it easier to spot clear trends). The figure contains four sub-figures that are color-coded by (i) the evolutionary state---whether a star is a red-giant branch (RGB) or a red-clump (RC) star---in the top, left, (ii) \numax\ in the top, right, (iii) \xh{Fe} in the bottom, left, and (iv) excess \Dnu\ in the bottom, right (see \eqnname\,\ref{eq:excessDnu}). 

In the top, left panel of \figurename\,\ref{fig:latent_colored}, we see that the RC stars in red are clearly clustered in most of the latent space dimensions. This is expected for the following two reasons. Firstly, the PSDs for RC stars look very different from those of RGB stars due to effects like mode-mixing (e.g., \citealt{2014A&A...572A..11G, 2014A&A...572L...5M}), such that one can distinguish RC from RGB with PSDs alone. The effect of these mixed modes can be more clearly seen when converting the PSD to an \'{e}chelle diagram---transforming the PSD to a two-dimensional image by stacking parts of the PSD separated by the large frequency separation \Dnu---such as those shown in \citet{2014ApJS..214...27M}. Secondly, pure spectroscopic separation of RC and RGB stars is also possible (e.g., \citealt{2014ApJ...790..127B, 2018ApJ...853...20H, 2018ApJ...858L...7T, 2022MNRAS.512.1710H}) and it is possible to obtain samples of RC stars with very high purity ($~5\%$ contamination) at low completeness. In the latent space, the RC cluster is not entirely separated from the RGB cluster, showing that there is a fundamental limit on the spectroscopic separability of the RC and the RGB for stars at the edges of both clusters. A clean seismic separation of RGB and RC requires the use of the period spacings expected from gravity modes \citep{2011Natur.471..608B}. Because we are not primarily interested in classifying RC/RGB stars, we do not pursue that here, but a better spectroscopic classification may be possible using a similar encoder-decoder approach applied to period spacings.

The latent space color-coded by \numax\ in the top, right panel of \figurename\,\ref{fig:latent_colored} shows a  smooth color gradient in some of the latent space dimensions. This is not surprising, as we have already demonstrated that our model can reconstruct power spectra fairly well in \figurename\,\ref{fig:recon} as discussed in \secname\,\ref{subsec:psd_recon}. The parameter \numax\ scales with the acoustic cut-off frequency $\nu_\mathrm{ac}$, which in turn scales with surface parameters such as \teff\ and \logg\ \citep{2009MNRAS.400L..80S} that can easily be determined from APOGEE spectra. 

When color-coding the latent space by \xh{Fe} in the bottom, left panel of \figurename\,\ref{fig:latent_colored}, we find no strong trend, even though there are numerous \xh{Fe} lines in the APOGEE spectrum and \xh{Fe} is one of the easiest parameters to obtain from APOGEE spectra. There is information in the PSD that potentially correlates with metallicity. For example, \citet{2017A&A...605A...3C} demonstrates that the amplitude of the granulation activity is significantly affected by metallicity. However, we are not sensitive to this, because we explicitly remove the background level of the PSD as discussed \secname\,\ref{subsec:kepler}. If the model is working as expected, we should not see metallicity information in the latent space and this is exactly what we see in the bottom, left panel of \figurename\,\ref{fig:latent_colored}. We perform a more detailed assessment of the amount of abundance information in the latent space in \secname\,\ref{subsec:realresult} below.

The bottom, right panel of \figurename\,\ref{fig:latent_colored} color-codes the latent space by excess \Dnu, which is the basic seismic parameter that is most sensitive to stellar mass. Excess \Dnu\ manifests itself in the PSD as small shifts in the separations of the oscillation modes compared to the expectation from Equation \eqref{eq:excessDnu}. We see clear trends with excess \Dnu\ in the latent space, indicating that the encoder has learnt to extract seismic information related to mass and to predict the detailed location of the oscillation peaks, rather than simply generating visually good-looking power spectra.

To further understand the latent space, we show in \figurename\,\ref{fig:latent_jacob} the Jacobian of the fourth dimension z[3] of the latent space with respect to each pixel in the APOGEE spectra (i.e., how changes in each pixel in the APOGEE spectra affects the value of z[3]) for RGB stars with low \numax\ and high \numax. Regions like the hydrogen lines (the Brackett series) in the APOGEE spectral range are known to be sensitive to \logg\ and we see that z[3] for RGB stars with low \numax\ is sensitive to the hydrogen lines, while this is not the case for stars with high \numax. Overall though, there is no one region in the infrared spectrum that contains the seismic information encoded in the latent space, instead it appears to be spread throughout the APOGEE spectral range.

\begin{figure}
\centering
\renewcommand\arraystretch{1.5}
\setlength\tabcolsep{0pt}
\begin{tabular}{c >{\bfseries}r @{\hspace{0.7em}}c @{\hspace{0.4em}}c @{\hspace{0.7em}}l}
  \multirow{11}{*}{\rotatebox{90}{\parbox{3cm}{\bfseries\centering APOKASC2 \\ Evolutionary State}}} & 
    & \multicolumn{2}{c}{\bfseries Latent Space Prediction} & \\
  & & \bfseries RC & \bfseries RGB & \bfseries Total \\
  & \rotatebox{90}{RC} & \MyBox{True Positive}{1,733} & \MyBox{False Negative}{73} & 1,806 \\[2.9em]
  & \rotatebox{90}{RGB} & \MyBox{False Positive}{222} & \MyBox{True Negative}{2,683} & 2,905 \\
  & Total & 1,955 & 2,756 & 4,711
\end{tabular}
\caption{Latent-space RC-vs-RGB classification. This figure shows the confusion matrix of a naive random forest model trained on the evolutionary state predicting whether a given APOGEE star is a RC or RGB star based on the latent space only. The model performs very well, only misclassifying $\sim 10\%$ of the stars. The RC misclassification rate (RC being misclassified as RGB) is $\sim 4\%$ while the RGB misclassification rate is $\sim 8\%$.} 
\label{fig:rc_confusion}
\end{figure}

\subsection{Abundances, seismic parameters and ages}\label{subsec:realresult}

As we discussed in \secname\,\ref{subsec:rf}, we use a modified version of the random forest method to map from the latent space to physical parameters such as abundances and stellar ages. Our primary goal is to determine ages from the latent space, but we also train random-forest regressors to determine abundances, seismic parameters, and the evolutionary state to better understand the information content of the latent space. We train separate random forest models for all of these cases.

\begin{description}
    \item \textit{Stellar parameters and abundances:} To quantitatively determine the abundance information content of the latent space, we train different random forest regressors to predict \xh{Fe}, \alpham, \xh{N}, and \cn\ from the latent space. We compare these predictions to the APOGEE ASPCAP values in \figurename\,\ref{fig:latent_aspcap}. We see in particular that the latent space is entirely unable to predict \alpham, as the latent-space prediction is simply the mean of the sample regardless of the true \alpham. Thus, the latent space has no information about \alpham. This in turn means that when we use the latent space to determine ages, these ages are entirely uninformed by \alpham. The \xh{Fe} and \xh{N} abundances show a small trend that still falls short of the 1-to-1 trend and the information in the latent space about these abundances is small. While there is only a weak trend in \xh{N} (and similarly, in \xh{C}), the latent space does appear to be able to predict \cn. This is consistent with the fact that the \cn\ ratio is a good mass proxy for giants \citep{2015MNRAS.453.1855M,2016MNRAS.456.3655M}, because mass correlates with the seismic parameters. However, the precision to which the latent space can predict \cn\ is still much worse than the precision to which it can be measured from the APOGEE spectra directly ($\sim 0.05$ dex). Thus, the \cn\ prediction here is not limited by how well one can determine \cn\ directly, but is instead limited by the fact that the encoder-decoder did not actually learn \cn\ directly. Rather, it learns other seismic parameters that correlate with \cn. In addition to the abundances, we check how well we can recover \teff\ from the latent space alone as \teff\ is needed by traditional asteroseismic age determination. We recover \teff\ from the latent space at a precision of $\sim 110$ K, similar to the precision to which one can recover \teff\ from simply predicting it from \logg\ for giants. So the latent space does not contain much information on \teff.
    
    \item \textit{Global seismic parameters:} We check how accurately we can predict the global seismic parameters \numax\, \Dnu\, and excess \Dnu\ from the latent space. A comparison between the seismic parameters taken from the APOKASC catalog and those predicted from the latent space is shown in \figurename\,\ref{fig:latent_seismic}. It is clear that in all three cases we can predict the global seismic parameters to high fidelity from the latent space.  Comparing to previous work on getting data-driven seismic parameters from PSDs, our work here has a comparable accuracy, although the nature of the previous work is very different. For example, \citet{2018ApJ...866...15N} predicts \numax\ and \Dnu\ at an accuracy level of $\sim15\%$ from the auto-correlation function (ACF) of the light curve using a very simple but interpretable polynomial model. \citet{2018ApJ...859...64H} develops a ``quick-look" deep learning method using convolutional neural networks to detect the presence of solar-like oscillations from 2D images of PSDs and then estimate \numax\ at a level of $\sim 5\%$. It is worth noting that converting a PSD to a low-resolution 2D image results in the loss of positional information of the oscillation modes, thus making the performance worse than it could in principle be. Hence, it is not surprising that our seismic parameter predictions from spectra are comparable to those from previous data-driven methods applied to PSDs. Comparing to the values from APOKASC, our predictions for \numax\ and \Dnu\ are at the accuracy level of $\sim 10\%$ and $\sim 7\%$ using the latent space while APOKASC's \numax\ and \Dnu\ precisions are at $\sim 0.02\%$ and $\sim 0.17\%$ respectively. This shows that the PSDs contains far more precise seismic information than can be obtained from the APOGEE spectra.

    \item \textit{Evolutionary state classification:} We saw in \figurename~\ref{fig:latent_colored} that RC and RGB stars cluster separately in the latent space, which means that we should be able to classify giants as RC or RGB stars using the latent space. Thus, we train a naive random forest classifier to perform this classification using the latent space. The confusion matrix of the resulting classification is shown in \figurename\,\ref{fig:rc_confusion}. Unlike works such as \citet{2014ApJ...790..127B} and \citet{2018ApJ...858L...7T}, our classifier does not aim at high purity at the expense of completeness, so it is expected that our purity is lower, while our completeness should be higher. Overall, we are able to obtain good classification results based on the latent space, only misclassifying $\sim 10\%$ of the stars. We see that our classifier is indeed able to obtain high completeness (only $\sim 4\%$ of RC stars are misclassified as RGB stars) at the expense of somewhat higher contamination  ($\sim 11\%$) than obtained in studies that focus on purity. We can improve the classification's performance along all axes by $\sim 1$ to $2\%$ by augmenting the latent space with \teff\ and \xh{Fe}.

    \begin{figure}
\centering
\includegraphics[width=0.475\textwidth]{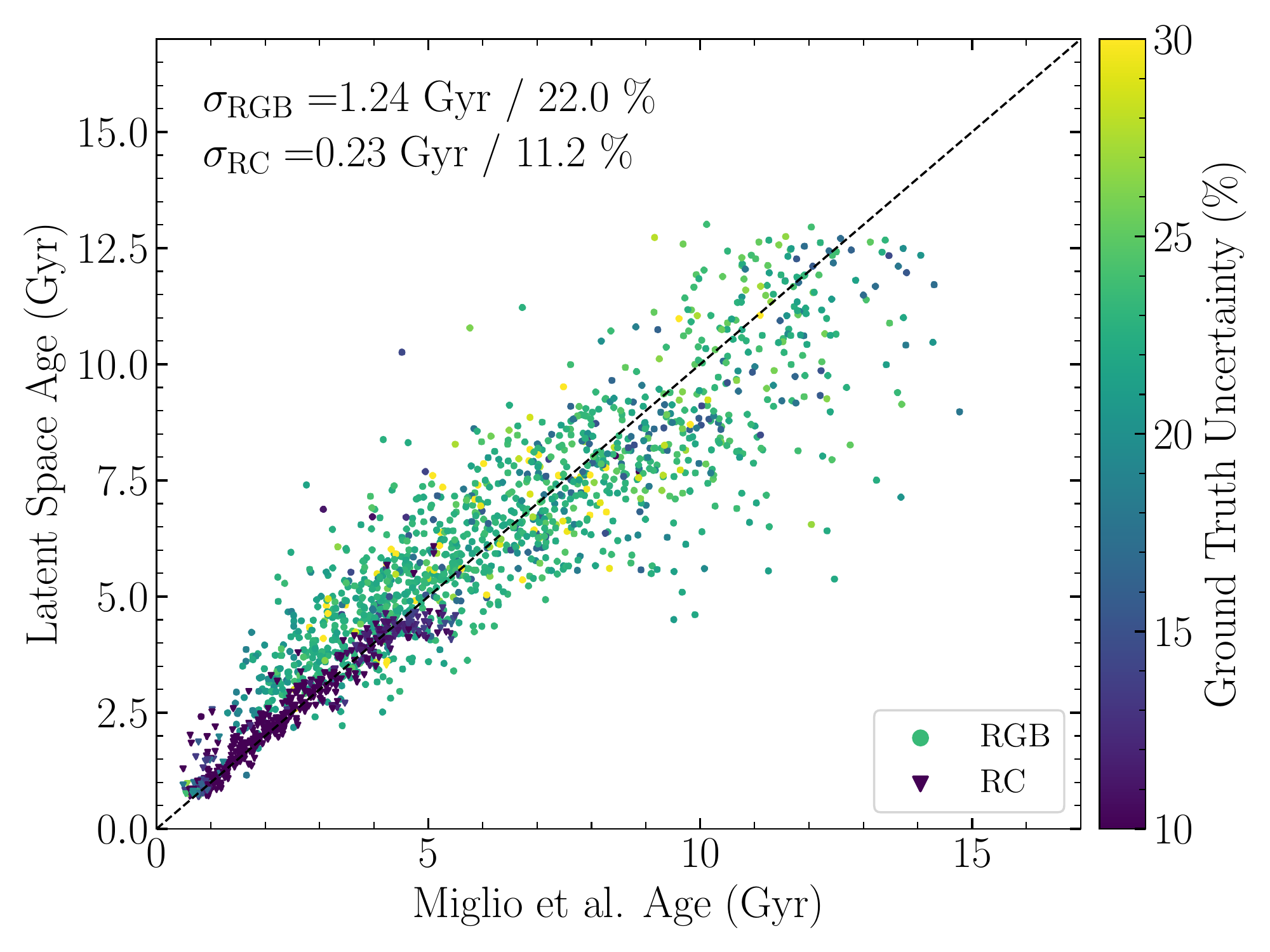}
\caption{Latent-space age predictions. This figure shows the age prediction for stars based on the latent space augmented by \teff\ and \xh{Fe} compared to the ground-truth value from \citet{2021A&A...645A..85M} for stars in the test set. Points are colored by the uncertainty in the  \citet{2021A&A...645A..85M} age with the marker shape indicating whether a star is an RC (triangle) or RGB (circle) star. The model is only trained on $\sim 1,200$ stars selected randomly from \citet{2021A&A...645A..85M}. Stars with very low ground truth uncertainty around $0-5$ Gyr old are RC stars and we are able to predict these ages to high precision. The standard deviation around the one-to-one line for RC and RGB stars separately are given in the top-left corner and are $\sim 11\%$ and $\sim 22\%$ respectively.}
\label{fig:age_test}
\end{figure}
    
    \item \textit{Ages:} Finally, we check how well we can predict age, which is the primary goal of this study. To obtain our fiducial age determinations, we augment the latent space with \teff\ and \xh{Fe}, because these are the spectroscopic parameters that a traditional asteroseismic age determination needs in addition to quantities derived from the light curves (i.e., \numax\ and \Dnu) and we have shown above that they are not available in the latent space. In \figurename\,\ref{fig:age_test}, we compare the ages that we obtain in this way with the ages from \citet{2021A&A...645A..85M}. We see that we are able to predict ages well over the entire age range of the sample: the predicted ages cluster around the one-to-one line. The overall bias is $\sim 3\%$ with an overall dispersion of $\sim 22\%$. However, the accuracy of our age predictions depend strongly on the evolutionary state: for RC stars, which are generally younger than RGB stars (ages typically in the range $0-5$ Gyr), have highly accurate age predictions with a scatter of only $\approx 11\%$. For RGB stars, on the other hand, the age accuracy is worse with a scatter of $\approx 22\%$. Nevertheless, this number represents a clear improvement to other data-driven spectroscopic ages including those from \citet{2019MNRAS.489..176M}, which uses a Bayesian neural network that is directly trained on APOKASC ages to obtain an age precision of $\sim 30\%$, and those from  \citet{2022MNRAS.512.2890L}, which uses the \texttt{Cannon} \citep{2015ApJ...808...16N, 2016arXiv160303040C}. Compared to \citet{2019MNRAS.489..176M}, we also do not observe any plateauing at $\sim 8$ Gyr, but we are instead able to obtain precise ages for old stars. Additionally, unlike the previous methods, we have demonstrated above that our ages are not informed by information comping from abundance ratios like \alpham, because this information is not contained in the latent space. We also obtain uncertainties on our predicted ages from the random forest regressor. To check the quality of these uncertainties, we compute the distribution of $ (\hat{y}-y) \Big/ \sqrt{{\sigma_{y}}^2 + {\sigma_{\hat{y}}}^2}$ where $y$ is the ground truth age, $\hat{y}$ is the model age, and $\sigma_{y}$ and $\sigma_{\hat{y}}$ are their respective uncertainties. If the uncertainties are correct, this distribution should be a standard normal distribution. Instead, we find that the distribution is normal with a width $\sim 0.75$, indicating that we are overestimating the uncertainty in our predicted ages. We provide further discussion of the predicted latent-space ages in \secname\,\ref{subsec:latentage} below.

\end{description}

\begin{figure}
\centering
\includegraphics[width=0.475\textwidth]{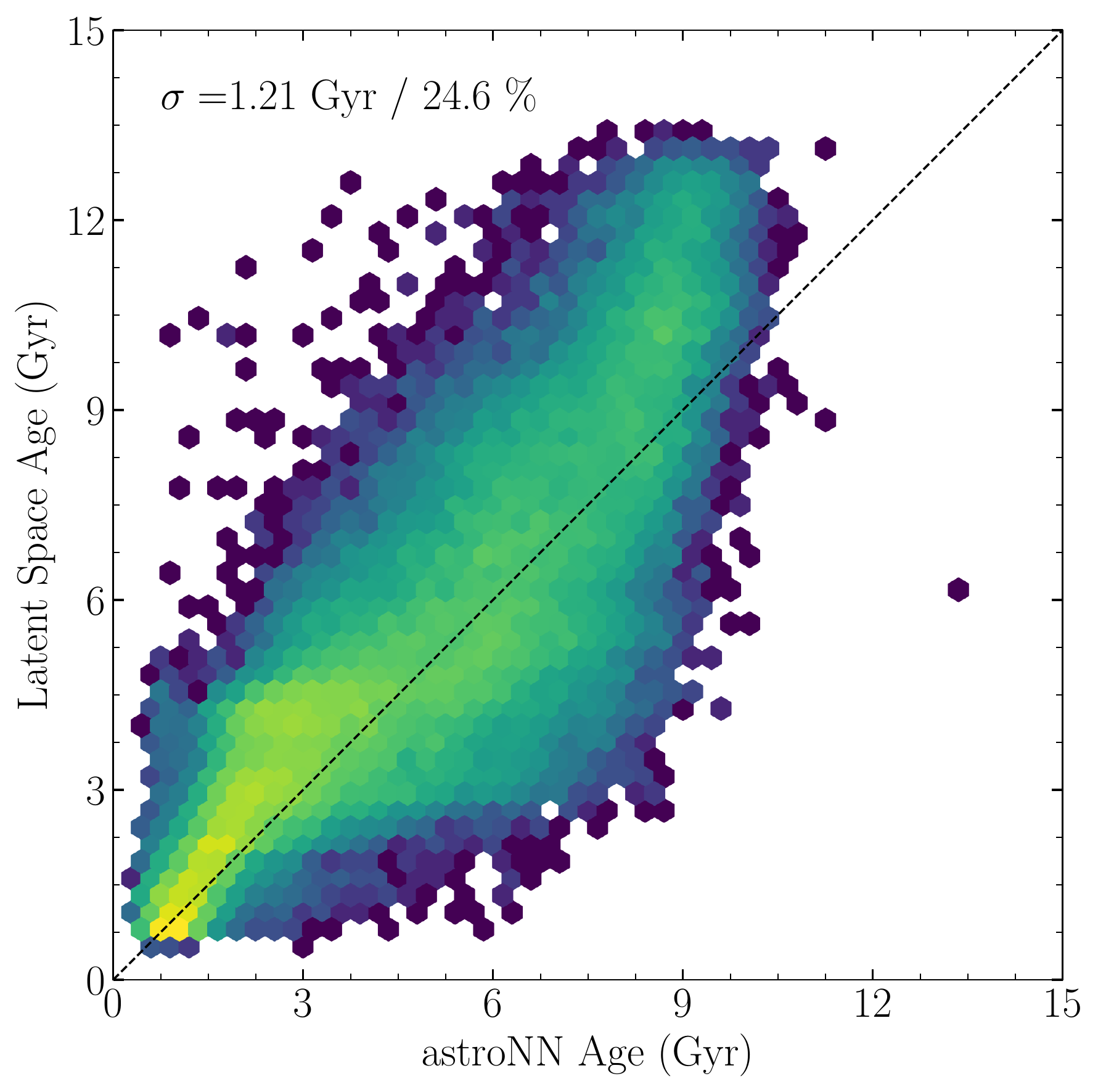}
\caption{Latent-space ages compared to spectroscopic ages from astroNN \citep{2019MNRAS.489..176M}. This figure shows the logarithmic density of a comparison of the latent-space ages from this work to the spectroscopic ages obtained using a Bayesian neural network trained directly on APOGEE spectra without physical constraints against using abundance information. The overall scatter between these ages predictions is $\sim 25\%$. Unlike the previous ages, the latent-space ages do not plateau around $8\,\mathrm{Gyr}$.}
\label{fig:astronn_vs_latent}
\end{figure}

\begin{figure*}
\centering
\includegraphics[width=0.95\textwidth]{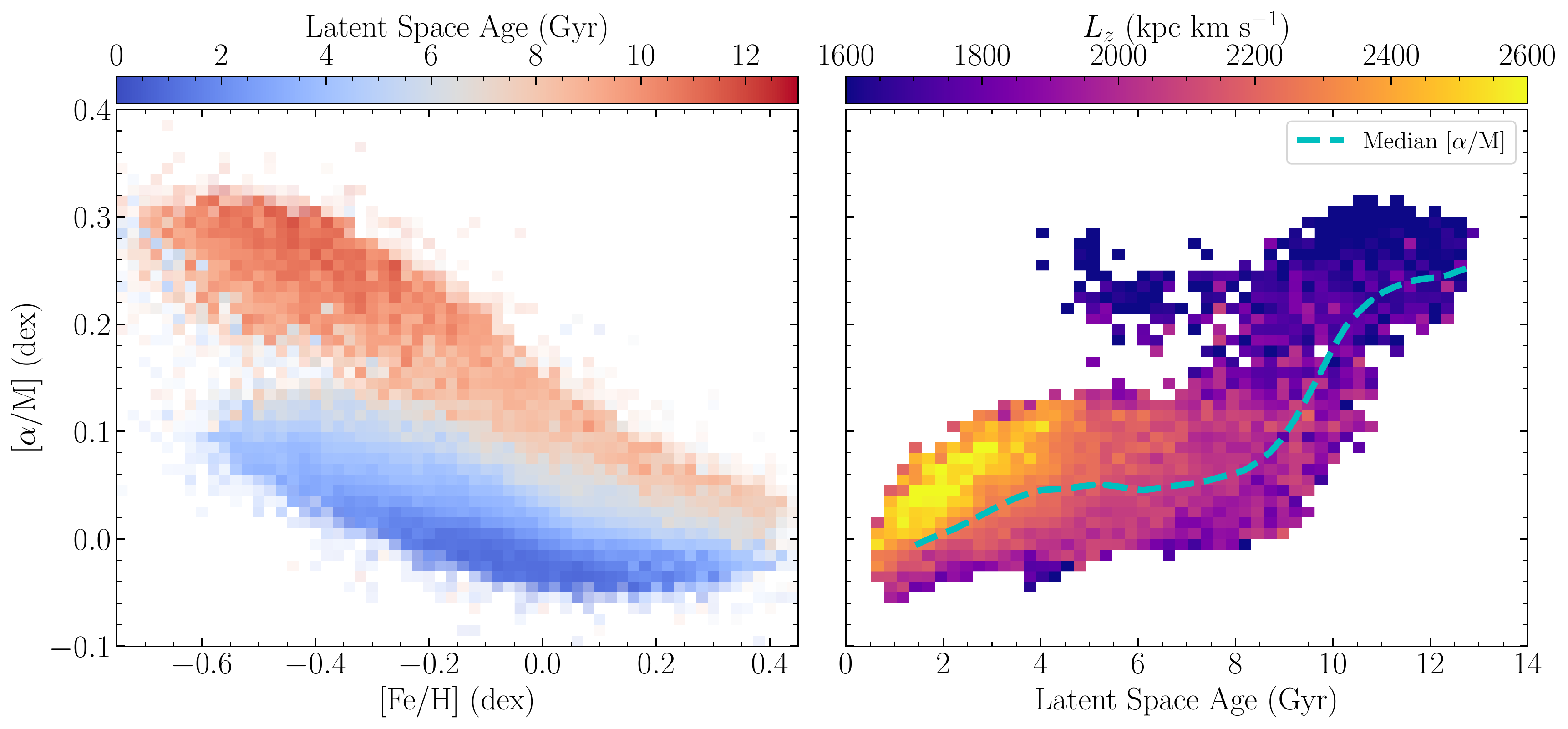}
\caption{Ages inferred from the latent space model applied to the whole APOGEE DR17 data set. After various quality cuts of spectral signal-to-noise ratio $\texttt{SNREV}>30$, latent space age $\sigma_\mathrm{age}< 50\%$, $2.5<\logg<3.3$, which is the range covered by the training set, as well as cutting on the \texttt{STARFLAG} and \texttt{ASPCAPFLAG} flags, $\sim 56,000$ giants are included in both panels. The left panel displays the \xh{Fe}-\alpham\ bimodality colored by average latent space age in each bin. The right panel shows the latent space age vs. \alpham\ colored by average angular momentum $L_z$ in each bin, with a cyan dashed line representing the median age-\alpham\ relation. Both panels clearly demonstrate that high \alpham\ sequence stars are significantly older than low \alpham\ sequence stars. In the right panel, there is a very small population of young ($\lesssim 6$ Gyr) \alpham-enriched stars that is not presented in the training set (see \figurename\,\ref{fig:miglio_data}) and that was previously observed using asteroseismic ages, but not usually in data-driven spectroscopic ages.}
\label{fig:dr17_age}
\end{figure*}

\begin{figure}
\centering
\includegraphics[width=0.475\textwidth]{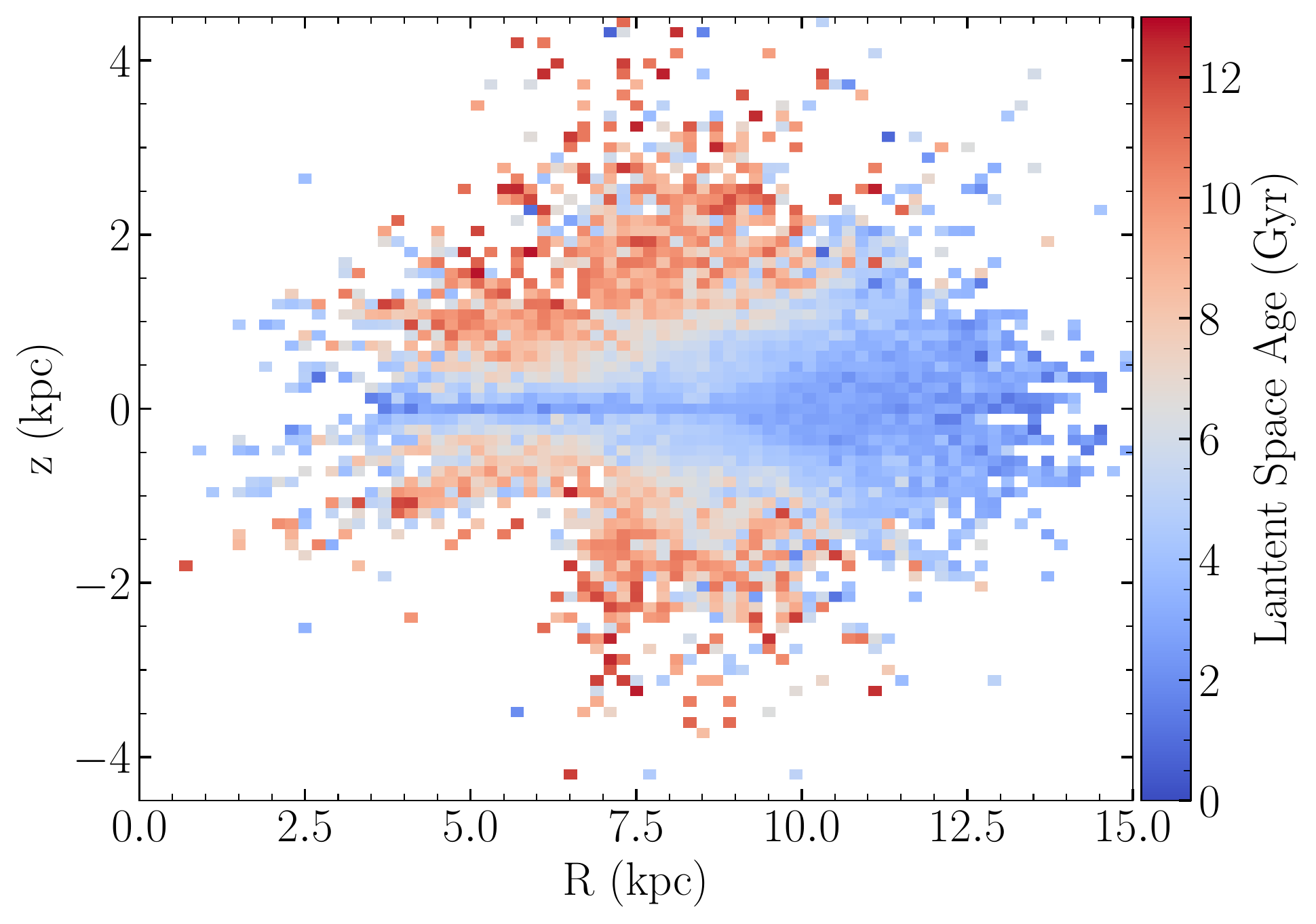}
\caption{Spatial distribution of stars in APOGEE DR17 in Galactocentric radius $R$ and vertical height $z$ colored by median latent space age. The spatial distribution clearly shows the vertical flaring in age of the disk with radius.}
\label{fig:z_age}
\end{figure}

\begin{figure*}
\centering
\includegraphics[width=0.95\textwidth]{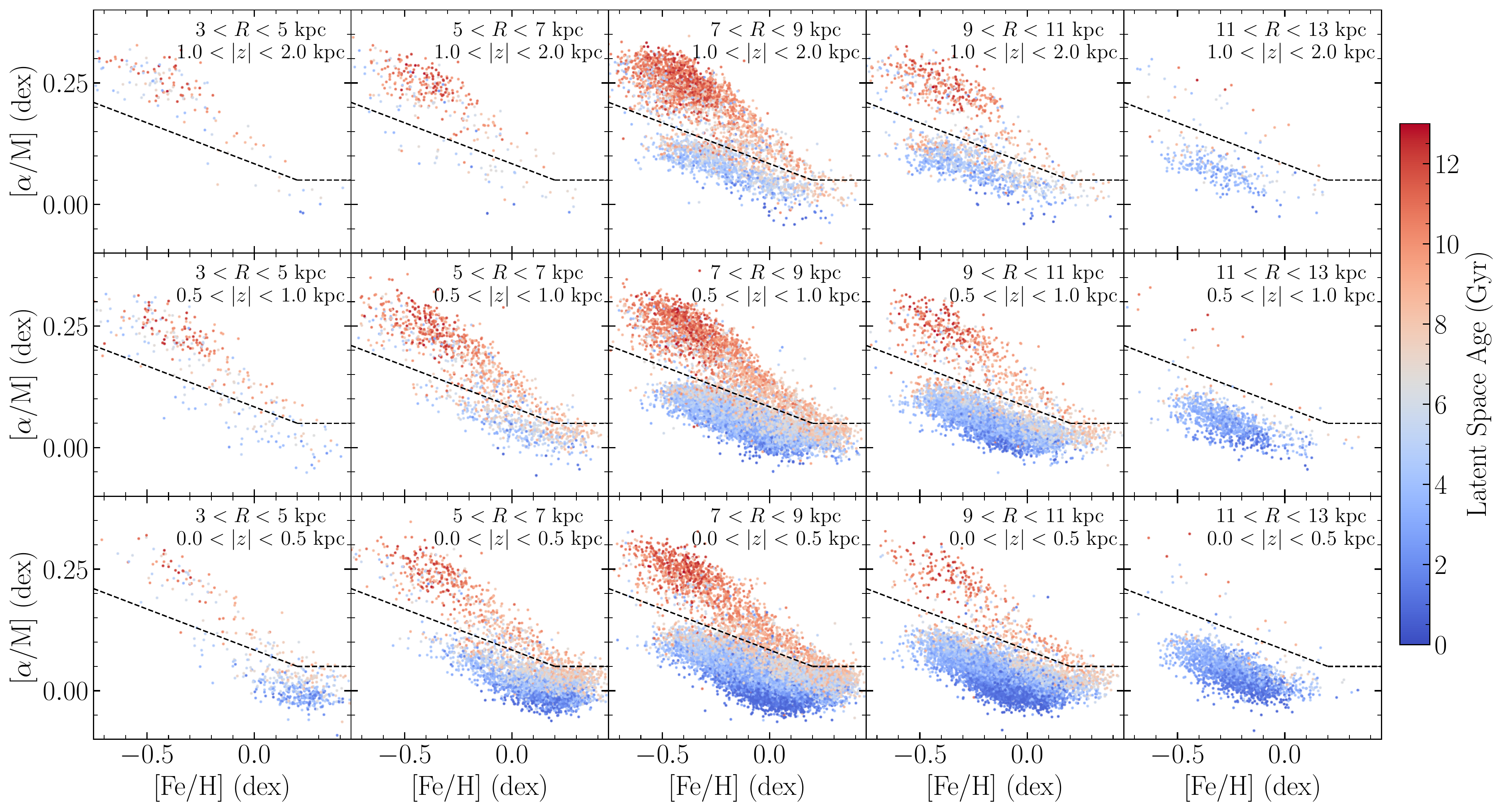}
\caption{The age-\xh{Fe}-\alpham\ distribution of stars in the Milky Way disk. This figure---similar to \figurename\,4 in \citet{2015ApJ...808..132H}---shows the \xh{Fe}-\alpham\ distribution color-coded by age of stars in spatial $R$-$z$ bins spanning $3\,\mathrm{kpc}<R<13\,\mathrm{kpc}$ in Galactic radius and $|z|<2\,\mathrm{kpc}$ in height from the Galactic midplane. The dashed line roughly divides the high- and low-\alpham\ sequences and is the same  for all subplots. It is clear that the high-\alpham\ sequence is older than the low-\alpham\ sequence with little age overlap between the two.}
\label{fig:alpha_Rz}
\end{figure*}

\section{The APOGEE DR17 latent-space age catalog and the age structure of the Milky Way disk}

\subsection{Age catalog}\label{sec:dr17_catalog}

We have applied our models to the whole APOGEE DR17 catalog to obtain latent space ages and produce a publicly available catalog. First, we compute the latent-space representation of APOGEE DR17 stars using the trained encoder-decoder for stars within the parameter space of the training set, that is, for stars with $1.5 < \logg< 3.6$ and $\texttt{SNREV}>30$, where \texttt{SNREV} is an alternative signal-to-noise measurement recommended by APOGEE that takes detector persistence issues in account. This allows $308,119$ stars to have their latent representation computed. To compute latent space ages, we further restrict the sample to stars with \teff\ and \xh{Fe} available as required by our pipeline, as well as lying in the same $2.5 < \logg< 3.6$ range as the  \citealt{2021A&A...645A..85M} training set used in this work. A data model and download link are available in \tablename\,\ref{table:data_model}. For science analyses, we recommend using stars with no \texttt{STARFLAG} and \texttt{ASPCAPFLAG} flag set as well as requiring a latent space age uncertainty less than $40\%$.

When compared to our previous work \citep{2019MNRAS.489..176M} of data-driven spectroscopic ages with neural networks as shown in \figurename\,\ref{fig:astronn_vs_latent}, the latent space ages from this work extend to significantly older ages for old stars, because unlike in our previous work, the latent space ages do not exhibit a plateau at around $8$ Gyr. Otherwise the two works are quite consistent with each other with a scatter of $25\%$. When plotting $\xh{Fe}-\alpham$ colored by latent space age as well as latent space age versus \alpham\ as shown in \figurename\,\ref{fig:dr17_age}, our latent space age shows the expected trends of low \alpham\ sequence stars being young and high \alpham\ sequence stars being old. The oldest stars in the low \alpham\ sequence are as old as the youngest high \alpham\ sequence stars. Although we have a number of young high \alpham\ stars, some of them can be removed by further restricting $\logg>2.55$ dex instead of $2.5$ dex to avoid the edge of training set parameter space and using our recommended $\sigma_\mathrm{age}< 40\%$ instead of the $50\%$ used in the figure for completeness.

For the purpose of studying spatial age--abundance trends in the Milky Way disk in the next subsection, we use the spectro-photometric distances from \citet{2019MNRAS.489.2079L} and convert from heliocentric to Galactocentric coordinates assuming $R_0=8.23$ kpc and $v_\odot=249.44\,\mathrm{km\ s}^{-1}$ \citep{2023MNRAS.519..948L} and $z_\odot=20.8$ pc \citep{2019MNRAS.482.1417B}. Orbital parameters are calculated using \texttt{galpy} \citep{2015ApJS..216...29B} using the standard \texttt{MWPotential2014} potential.

\subsection{The age structure of the Milky Way disk}\label{subsec:agetrend}

As a first application of our new age catalog, we present a brief investigation into spatial age-abundance trends in the Milky Way disk here. In \figurename\,\ref{fig:dr17_age}, the left panel shows the \xh{Fe}-\alpham\ distribution colored by mean latent space age in each bin. We see a clear separation between old high and young low \alpham\ sequences. The right panel shows the age-\alpham\ distribution colored by angular momentum $L_z$ with the cyan dashed line representing the median relation between age and \alpham; the \alpham\ scatter around this relation is $\approx 0.05$ dex for any age. The overall trend is a slowly increasing \alpham\ with age for the low \alpham\ sequence and a steeper trend when transitioning to the high \alpham\ sequence, similar to what is seen for local samples \citep{2013A&A...560A.109H}. The trend with angular momentum shows that the \alpham\--age trend is steeper in the high-angular momentum, outer disk than it is in the inner disk.

In the right panel of \figurename\,\ref{fig:dr17_age}, we also see that there are young, high \alpham\ stars with ages $\sim 6$ Gyr, similar to previous works (e.g., \citet{2015MNRAS.451.2230M, 2015A&A...576L..12C}), which are likely high \alpham\ stars with unusual elemental abundance ratios (e.g., \citealt{2019MNRAS.487.4343H}) that seem to be over-massive (and thus appear young) due to binary stellar evolution \citep{2021ApJ...922..145Z, 2022arXiv220711084J}. It is worth noting that when we adopt the definition of young, high \alpham\ stars with a flat cut of $\alpham >0.15$ dex and age younger than 6 Gyr used in previous literature (e.g., \citealt{2015MNRAS.451.2230M}), we find that $9\%$ of the high \alpham\ sequence population is young. This can be compared to $\approx 6\%$ in \citet{2015MNRAS.451.2230M} and \citet{2021ApJ...922..145Z}. Further restricting to $\logg>2.55$ dex, the fraction of young high \alpham\ stars approaches $7\%$. Alternatively, defining high \alpham\ as $\alpham>0.18$ dex, similar to values adopted in asteroseismic analyses, the fraction is $6\%$. These results are interesting, because there are no young high \alpham\ stars in our training set, yet we recover similar fractions of them as independent, previous analyses.

The spatial distribution of stars in Galactocentric radius $R$ and vertical height $z$ is shown in \figurename\,\ref{fig:z_age}. We clearly see the vertical flaring of the disk in age with radius. The outer disk is uniformly younger than $\approx 5\,\mathrm{Gyr}$. 

The age-\xh{Fe}-\alpham\ distribution of stars in Galactocentric radius $R$ and vertical height $|z|$ bins is displayed in \figurename\,\ref{fig:alpha_Rz}. The dashed black line that we use to separate the low and high \alpham\ stars is given by the combination of the following equations 
\begin{equation} \label{eq:alpha_cut}
\begin{split}
    \alpham &> -0.2211 \times \xh{Fe}+0.0442 \,\mathrm{dex},
    \\
    \alpham &> 0.05 \,\mathrm{dex}
\end{split}
\end{equation}
We see that the high \alpham\ sequence is old wherever it appears, with age declining from $\approx 12\,\mathrm{Gyr}$ at its low-metallicity end to $\approx 8\,\mathrm{Gyr}$ at its high-metallicity end; the smattering of young high \alpham\ stars is uniformly mixed up with these, again demonstrating that these are likely anomalously massive rather than anomalously young stars. The low \alpham\ stars are generally younger than $8\,\mathrm{Gyr}$. While the age--abundance trends seen in this figure are generally what has been found before, the precision of our ages for a large sample of stars sharpens the picture significantly compared to previous work.

\begin{figure*}
\centering
\includegraphics[width=0.95\textwidth]{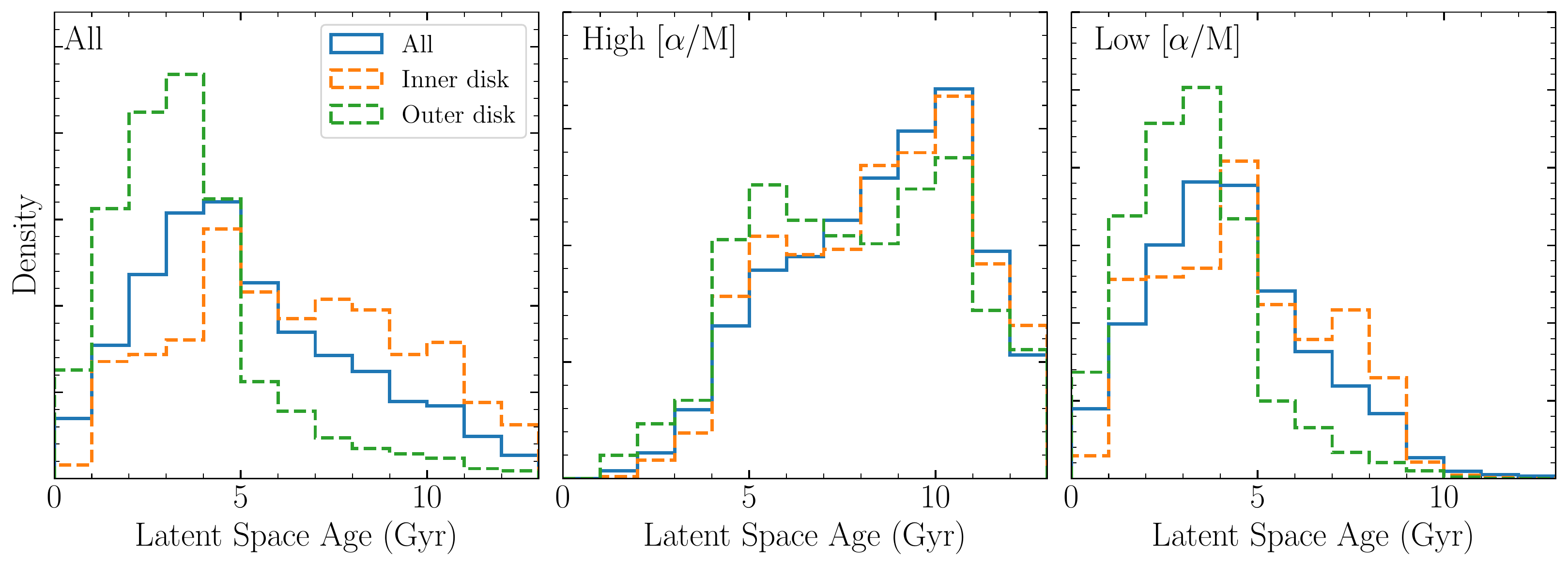}
\caption{Stellar age distributions in the Milky Way. Each panel in this figure displays the age distribution stars across the entire disk (`All') and those in the inner and outer disks, defined as  $3\,\mathrm{kpc}<R<6\,\mathrm{kpc}$ and $10\,\mathrm{kpc}<R<13\,\mathrm{kpc}$, respectively. The left panel shows all stars, regardless of their abundances, while the middle and right panels split the stars into high \alpham\ and low \alpham\ using the dashed line in \figurename\,\ref{fig:alpha_Rz}. The left panel demonstrates that outer disk stars are significantly younger than inner disk stars. The middle panel shows that high \alpham\ stars are old, but extend down to ages of $5\,\mathrm{Gyr}$.  The right panel demonstrates that low \alpham\ sequence stars are young across the disk and do not exceed $9\,\mathrm{Gyr}$ (accounting for uncertainties, the low \alpham\ upper age cut-off is $\approx 8,\mathrm{Gyr}$).}
\label{fig:age_alpha_hist}
\end{figure*}

A more detailed view of the age distribution and its overall radial and abundance trends is presented in \figurename\,\ref{fig:age_alpha_hist}. While the giant sample selected by our cuts does not sample age uniformly, calculations similar to those in Section 5 of \citet{2014ApJ...790..127B} show that the relative age bias of $\approx 2$ to $5\,\mathrm{Gyr}$ to $\approx 10\,\mathrm{Gyr}$ is a factor of $\approx 2.5$, with the age bias being relatively flat between 1 and $5\,\mathrm{Gyr}$ and then slowly decreasing towards larger ages. Given this, the overall age distribution in the left panel is indicative of a roughly uniform intrinsic age distribution, or a flat star-formation history. Also from the left panel, we see that the inner disk is more heavily weighted towards old ages, while the outer disk barely extends beyond $5\,\mathrm{Gyr}$ (see also the rightmost panels of \figurename\,\ref{fig:alpha_Rz}). The middle and right panels split these radial age trends by \alpham\ abundance, defining high and low \alpham\ sequences using the separation from Equation \eqref{eq:alpha_cut}. The age distribution of the high \alpham\ population is approximately the same regardless of radius and peaks at $\approx 10\,\mathrm{Gyr}$. Accounting for the age bias, there is a tail towards younger ages, but most of these are actually the likely anomalously-massive high \alpham\ stars discussed above. \figurename\,\ref{fig:alpha_Rz} demonstrates that the high \alpham\ sequence ends at $\approx 8\,\mathrm{Gyr}$. The low \alpham\ sequence has a clear radial age trend, with the inner disk being older than the outer disk. Regardless of radius, the low \alpham\ disk is younger than $\approx 9\,\mathrm{Gyr}$, or likely younger than $\approx 8\,\mathrm{Gyr}$ once we account for age errors (we, however, do not attempt a deconvolution of the age distribution in the first look here). This all indicates that the high and low \alpham\ sequences formed at different times, with the high \alpham\ sequence corresponding to the first $\approx 4\,\mathrm{Gyr}$ of the disk's existence and the low \alpham\ sequence corresponding to the last $\approx 8\,\mathrm{Gyr}$. There does not appear to be a large hiatus between the two, although properly understanding the transition would require proper deconvolution of the age distribution and a good understanding of the anomalous young high \alpham\ stars. A similar picture has emerged previously from observations near the Sun (e.g., \citealt{2013A&A...560A.109H,2020ApJ...897L..18B}).

\begin{figure*}
\centering
\includegraphics[width=0.95\textwidth]{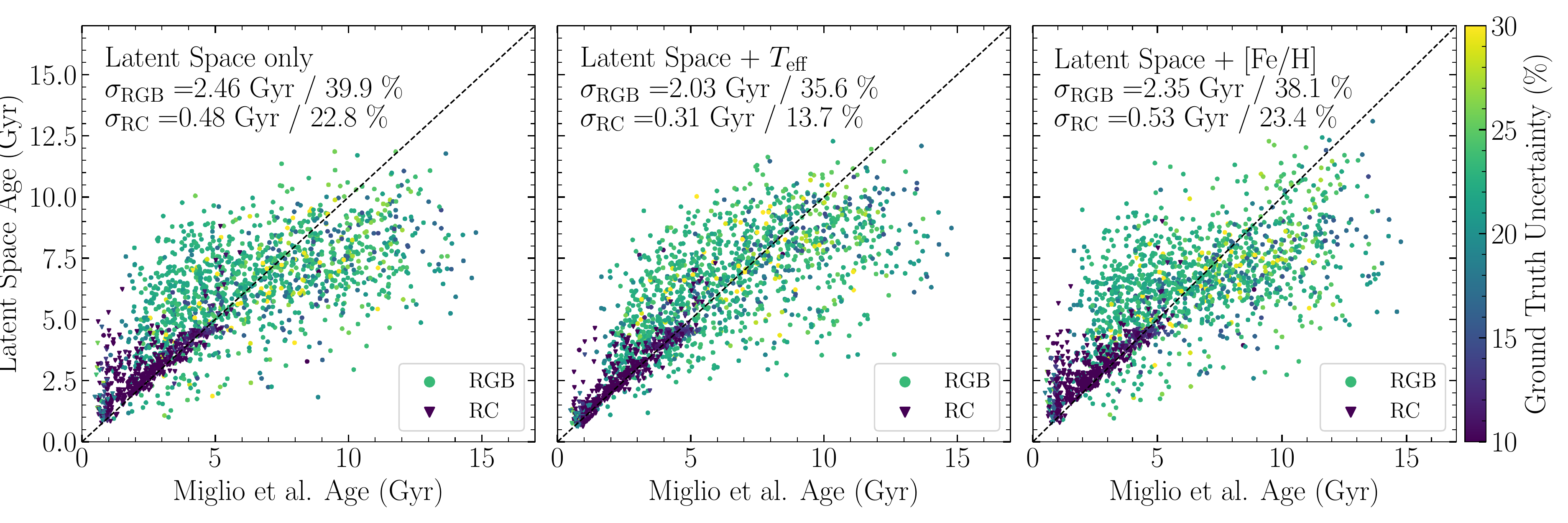}
\caption{Latent space age prediction using different combinations of parameters. The left panel shows the prediction based on the latent space only without any additional parameters. The middle panel displays the prediction based on the latent space augmented with \teff\ and the right panel's age is based on the latent space with \xh{Fe}. These can be compared to our fiducial latent space ages, which are based on the latent space augmented with both \teff\ and \xh{Fe} and for which this comparison is shown in \figurename\,\ref{fig:age_test}. In all three cases, the age prediction is significantly worse that in our fiducial model, with much larger scatter compared to the ground-truth ages and with a plateau arising at old ages ($\gtrsim 10$ Gyr).}
\label{fig:latentage_3}
\end{figure*}

\begin{figure}
\centering
\includegraphics[width=0.475\textwidth]{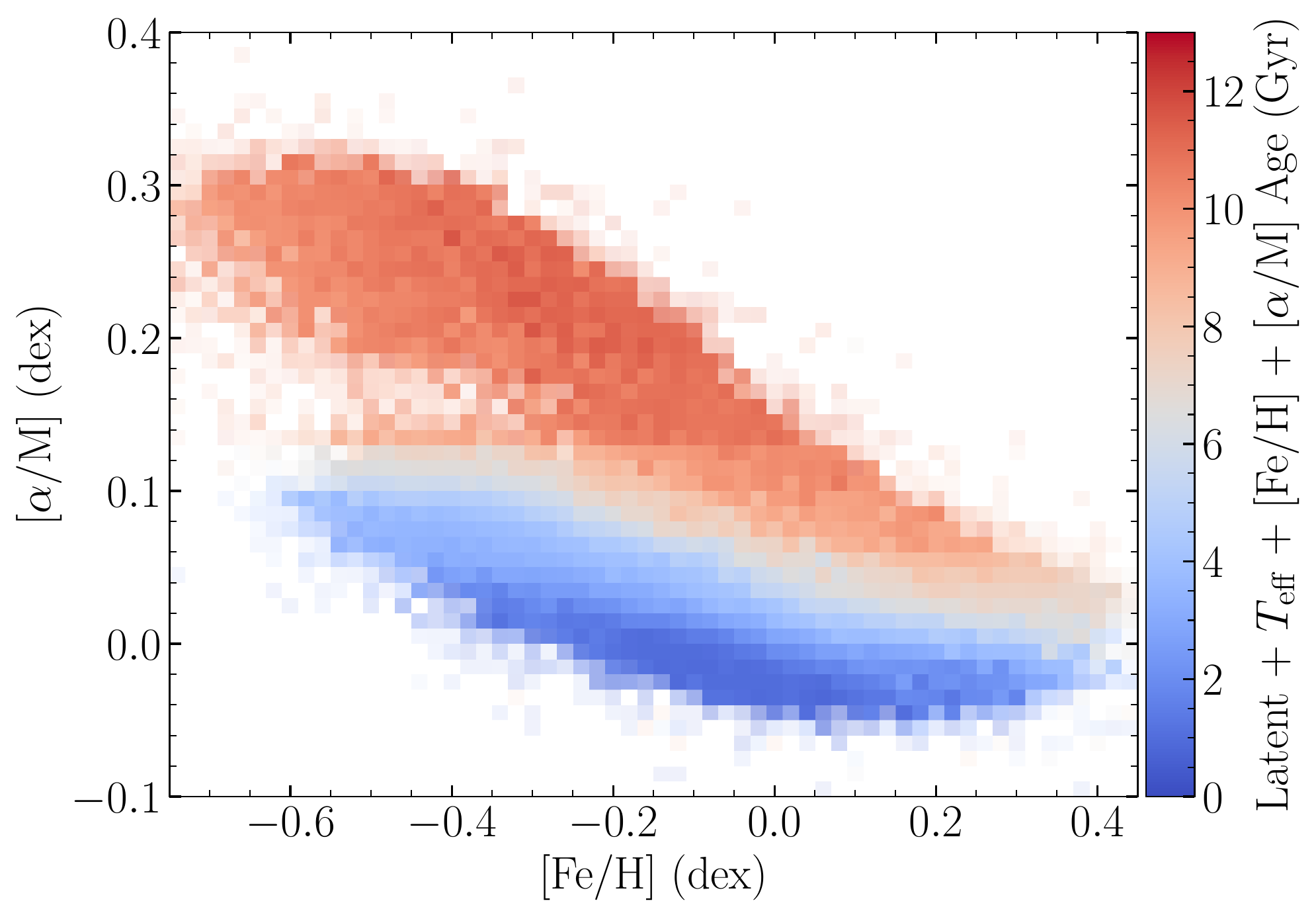}
\caption{The impact of including \alpham\ in the latent-space age model. This figure is similar to the left panel of \figurename\,\ref{fig:dr17_age}, but the latent space age here is obtained with model using the combination of the latent space, \teff, \xh{Fe}, and \alpham. The additional \alpham\ on top of the other parameters smooths the \alpham\ bi-modality plot significantly compared to the left panel of \figurename\,\ref{fig:dr17_age}, with additionally artifacts appearing at $\alpham \sim 0.13$ dex because of the \alpham\ selection from our training set (see \figurename\,\ref{fig:miglio_data}). This demonstrates that including \alpham\ information in spectroscopic age determination strongly impacts the derived age--abundance trends.}
\label{fig:latent_with_aplha}
\end{figure}

\section{Discussion} \label{sec:discussion}

\subsection{Latent Space Ages}\label{subsec:latentage}

In the previous section and in \figurename\,\ref{fig:age_test}, we have demonstrated that we can obtain precise ages from the latent space augmented by \teff\ and \xh{Fe}. To understand how important each of these three ingredients is to the prediction of the age, we have tested a few other combinations of these ingredients. In \figurename\,\ref{fig:latentage_3}, we use a three separate models to predict age. The first model, shown in the left panel, uses the latent space only, the second model, in the middle, uses the latent space and \teff, and the third model, in the right panel, uses the latent space and \xh{Fe}. These three models allow us to assess the relative importance of the latent space and the augmented parameters in the age prediction. The age predictions in all three of these models are significantly worse than our fiducial latent-space age. Interestingly, the predicted age plateaus at $\sim 8$ Gyr similar to what we observed in \citet{2019MNRAS.489..176M} when directly predicting ages from APOGEE spectra. When only using the latent space, the age prediction still follows the one-to-one relationship relatively well, albeit with large scatter, indicating that the seismic information in the latent space is rich enough to provide a rough estimate of the age. While the scatter in the predicted ages decreases when adding \teff\ and \xh{Fe} separately, it is clear from comparing the middle and right panels to the left panel that the addition of these parameters separately does not qualitatively improve the age prediction over that from the latent space alone. Thus, adding both \teff\ and \xh{Fe} to the latent space is crucial to the good performance in 
\figurename\,\ref{fig:age_test}.

In \figurename\,\ref{fig:latentage_3}, it is clear that RC stars get good ages regardless of which combination of latent space, \teff, and \xh{Fe} we use. This is a direct consequence of the scaling relations combined with the fact that RC stars have a narrow range of absolute magnitudes. The mass scaling relation in \eqnname\,\eqref{eq:scaling} can be expressed in terms of the luminosity instead of \teff\ as shown in Equation (7) of \citet{2012MNRAS.419.2077M}. Because of the narrow luminosity spread of RC stars \citep{1998ApJ...494L.219P}, good ages for RC stars can thus be obtained with seismic data only assuming no significant mass loss. This means that we can obtain good RC ages just using the latent space.

That not using both \teff\ and \xh{Fe} degrades the age predictions is expected, because traditional asteroseismic age pipelines require at least the following parameters to predict the age: seismic information (typically in the form of \numax\ and \Dnu), \teff, and \xh{Fe}. This is because we can obtain the stellar mass using the seismic information and \teff\ (\eqnname\,\ref{eq:scaling}) and converting mass to age mostly relies on the \xh{Fe}-dependent main-sequence lifetime of the star. While we expect the latent space to contain all the seismic information that can be extracted from the APOGEE spectra, it does not contain \teff\ as we have discussed in \secname\,\ref{subsec:realresult} or \xh{Fe} as we have shown in \figurename\,\ref{fig:latent_colored}. Thus, it is not surprising that the age precision degrades when dropping one of \teff\ and \xh{Fe}.

Finally, we investigate how the latent space ages change when we also include \alpham\ in the prediction. \figurename\,\ref{fig:latent_with_aplha} shows the age prediction for a model that includes \alpham\ directly in addition to the latent space, \teff, and \xh{Fe}. Although including \alpham\ does not cause age to correlate much more strongly with \alpham, the inclusion of \alpham\ does significantly smooth the age trend in the \xh{Fe}-\alpham\ space while only improving the performance of our model by about $2 \%$. Therefore, even thought \alpham\ is also a parameter sometimes used by stellar models to get ages, we choose not to include \alpham\ to avoid the adverse effects that result from its inclusion.

Our current latent space age model suffer from the limitations of the age, because the model can only be applied to new stars in the same parameter space as that covered by the training sets of the encoder-decoder and the age. In this case, the most limiting factor is the narrow $2.5<\logg<3.6$ range of the age training set.

\subsection{Application with TESS and Beyond}\label{subsec:tess}

The NASA Transiting Exoplanet Survey Satellite (\tess) mission is an all-sky survey that allows for the detection of solar-like oscillations in at least an order of magnitude more giants than \kepler\,\citep{2020ApJ...889L..34S, 2021MNRAS.502.1947M, 2021ApJ...919..131H, 2022AJ....164..135H}. Compared to \kepler\, or even the K2 mission \citep{2014PASP..126..398H}, \tess\ light curves are typically much shorter and less precise, making it more difficult to do precise asteroseismology.

We can exploit the flexibility of the encoder-decoder model to include \tess\ light curves. To train our method, we require pairs of APOGEE spectra and light curves that cover a larger parameter space than that spanned by the training set of the ages (i.e., we cannot train the latent space $\rightarrow$ age regression with stars outside of the parameter space of the training set of the encoder-decoder; the parameter space spanned by the training set of the encoder-decoder is much larger than that of the latent space age). Including light curves from \tess\  can help enlarge the parameter space with stars with a wider range of chemical abundance patterns from different parts of the Galaxy, especially stars with low metallicity ($\lesssim -1.5$ dex) that are not abundant in the \kepler\ field.

Data-driven models, and in particular deep neural networks, are susceptible to small changes in the data. For example, a trained neural network to predict abundances from spectra cannot typically be applied to data taken by different instruments, even if they cover the same wavelength range. Also, it is difficult to train a model on synthetic data (e.g., synthetic spectra) and then apply it to observational data (e.g., observed spectra); often this type of analysis will show a ``synthetic gap'' between the synthetic and true data that adversely affects the performance of  the method (e.g., \citealt{2018MNRAS.475.2978F}). In the application at hand here, \tess\ and \kepler\ have different instrumental noise properties, different photometric precision, different observing cadences, etc. Naively combining \tess\ and \kepler\ light curves will likely make our model worse (see, e.g., \citealt{2021ApJ...919..131H} for a discussion of the difference between \tess\ and \kepler\ for machine learning models). Making use of better algorithms to generate the PSD (e.g., the multi-taper algorithm of \citealt{2022arXiv220915027P}) may result in more uniform PSDs derived from observational data. We could also use different styles of transfer learning that can map data taken with different instruments onto a uniform scale (e.g., \citealt{2021ApJ...906..130O}. For example, we could employ another encoder-decoder to map \tess\ PSDs to \kepler\ PSDs trained on overlapping observations between \tess\ and \kepler. Future missions like PLATO \citep{2014ExA....38..249R} and HAYDN \citep{2021ExA....51..963M} will also have their own photometric precisions, baselines, and cadences and would therefore also benefit from such an approach..

\subsection{Prospects for SDSS-V}\label{subsec:sdssv}

The Milky-Way Mapper (MWM) in SDSS-V is an ongoing panoptic survey similar to the APOGEE survey that uses the same spectrograph, but combines it with a robotic focal plane system (FPS) with 300 robotic fiber positioners allowing flexible, high-cadence targeting and obtaining high target densities in a small field of view \citep{2020SPIE11447E..81P}. The Galactic Genesis Survey (GGS) program within MWM will produce a densely sampled panoptic spectroscopic stellar map covering a large volume of the Milky Way by targeting very luminous giants with low \logg\ (typically lower \logg\ than APOGEE observations). As we have discussed in \figurename\,\ref{fig:dr17_age} and \secname\,\ref{subsec:latentage}, currently the latent space age model only applies to the parameter space covered by the ages of \citet{2021A&A...645A..85M}, with the major limitation coming from the narrow range of \logg\ present in that sample. Because many stars in GGS have lower \logg\ than this, our model will not be directly applicable to GGS/MWM observations. In particular, the \logg\ limitation makes it difficult to reach the bulge with current APOGEE-like observations (see \figurename\,\ref{fig:z_age}).

The minimum useful frequency we can obtain from the Lomb-Scargle PSD of \kepler\ light curves is about $3\,\uHz$ and this minimum has been adopted by many previous papers (e.g., \citealt{2018ApJ...866...15N, 2018ApJ...859...64H}). While we might be able to determine \numax\ down to this minimum frequency, realistically we can only obtain precise determinations of additional global seismic parameters as \Dnu\ at $\Dnu \gtrsim 7\,\uHz$. This cut-off corresponds to giants with $\logg \sim 1.8$ dex using standard scaling relations. Aside from the fact that the estimating mass and age using the empirical scaling relations might break down at very low \numax, the minimum frequency sets a lower \logg\ limit on applying our encoder-decoder methodology to luminous giant stars. The \logg\ limitation along with the lack of low-metallicity stars in the \kepler\ field currently makes it difficult to train and apply our method to obtain age estimates for interesting objects such as the Gaia-Enceladus merger remnant \citep{2018Natur.563...85H}, which shows up in APOGEE in significant numbers only at low \logg\ and low metallicity.

\subsection{Using alternative mass measurements to train}\label{subsec:altermass}

The two-component nature of our methodology, where we first extract seismic information that contains mass information into the latent space using the encoder-decoder and we then map the latent space to age, means that we have flexibility in how we train the latent space $\rightarrow$ age regression. In the second step, we do not need to use ages derived from the PSDs used to train the encoder-decoder, but we could instead use ages obtained from other parts of the sky (e.g., the \tess\ continuous viewing zone), as long as the age sample has similar stellar populations as the sample used to train the encoder-decoder, because we need the latent-space representation of the age sample. We do not even need to use seismic ages at all in the second step, although we do so here because currently asteroseismology is the only method that provides precise ages for large-ish samples of stars. As we have shown in this work, we only need about about a thousand accurate age measurements as the training set for the latent space age.

Because we can convert mass to age relatively precisely along the giant branch, we could use alternative mass determinations to train the latent space age. For example, we could use stars in a cluster with masses determined using stellar evolution models and we could even require that they return the same age for all stars in the cluster. We could also make use of masses determined for eclipsing binaries using Kepler's laws, as these allow mass determinations at the few percent level compared with the $\sim 8\%$ masses for giants from asteroseismology. That these methods can obtain such highly accurate masses and ages has been demonstrated by \citet{2012A&A...543A.106B} 
 in estimating a cluster's age using binary members as well as by \citet{2013ApJ...767...82G} and \citet{2018MNRAS.476.3729B} in estimating age with eclipsing binary systems.

As long as we can get APOGEE spectra for giants with such alternative mass measurements , we can include them in the training to go from the latent space to age, because we can determine their latent space parameters using the encoder (we do not need the decoder). Precise astrometry from \gaia\ can potentially detect tens of thousands of resolved wide binaries from which precise masses can be determined (e.g., \citealt{2017MNRAS.472..675A, 2017A&A...602A.110K, 2017A&A...606A..92M, 2021MNRAS.506.2269E, 2022arXiv220605595G}) for an all-sky sample.

\section{Conclusions}\label{sec:conclusion}

We are living in an era of abundance asteroseismic and spectroscopic data. Surveys such as \tess\ and the upcoming PLATO mission \citep{2014ExA....38..249R} will allow for large sets of ages for giant stars to be determined, which are essential for Galactic archaeology. At the same time, even larger spectroscopic surveys are ongoing, like SDSS-V and the upcoming 4MOST \citep{2012SPIE.8446E..0TD} survey, which collect stellar spectra densely-sampled across the sky and covering large parts of the Milky Way disk and halo. Thus, being able to determine ages using spectroscopic data allows for detailed investigations into our Galaxy's formation and evolution.

In this paper, we have applied a well developed methodology in deep learning called a variational encoder-decoder to create a data-driven model that can determine more precise ages from APOGEE spectra compared to other data-driven methods by leveraging available asteroseismic data, provided that there is a large sample of spectrum--light-curve pairs (which do not require age determinations). We train a model on $\sim 10,000$ pairs of APOGEE spectra and \kepler\ light-curves to reduce the dimensionality of APOGEE spectra while simultaneously extracting mass and age information without contamination from abundance information beside \xh{Fe}. Reducing the dimensionality of APOGEE spectra in a latent space is crucial for being able to train an age model, because precise giant ages are rare and it is, therefore, difficult to train a complex model to infer spectroscopic age. We then trained a simple random forest model to determine ages from the latent space of the encoder-decoder model.

We showed that we are able to reduce the dimensionality of APOGEE spectra to just five dimensions for the purpose of reconstructing the relevant information in the light curve's PSD. The decoder is able to reconstruct the PSD in the region where pressure modes are located. From the resulting latent space, we are able to infer ages precise to $\sim 22\%$ by training only with $\sim 1,200$ stars with good ages using the latent space augmented by \teff\ and \xh{Fe}; for red clump stars we approach a precision of $\sim 10\%$. We have applied our methods to the whole APOGEE DR17 catalog for stars that fall within the parameter set of the training data. The resulting latent space ages are overall consistent with the data-driven spectroscopic ages from \citet{2019MNRAS.489..176M}, except that old stars are much older (there is no plateau at the old end as in previous work) and the \alpham\ rich stars are significantly older than the oldest \alpham\ poor stars. Because our latent space does not include information on \alpham\ and only weak information about other abundance ratios, our latent space ages are independent of abundance ratios, yet we will obtain precise ages. 

Using the APOGEE DR17 age catalog that we create in this paper, we have mapped the age-abundance distribution across the radial range spanned by the Galactic disk. We find that the high \alpham\ sequence has the same age distribution at any radius, extending from ages of $\approx 12\,\mathrm{Gyr}$ to $\approx 8\,\mathrm{Gyr}$; at younger ages, we find a small fraction of young, high \alpham\ stars similar to what has previously been found. The low \alpham\ disk is younger than $\approx 8\,\mathrm{Gyr}$ everywhere, with a radial gradient in the oldest low \alpham\ stars: the outer disk ($R \gtrsim 10\,\mathrm{kpc}$) is almost entirely low \alpham\ and younger than $\approx 5\,\mathrm{Gyr}$. The high and low \alpham\ disks appear to be temporally separated, with the high \alpham\ disk representing the early evolution of the disk that transition to the later low \alpham\ evolution $\approx 8\,\mathrm{Gyr}$ ago. 

The PSD reconstruction provided by our encoder-decoder has interesting applications of its own. For example, it can be used as a ``sanity check" for the light curve, because we can quickly check whether the directly observed PSD deviates greatly from that reconstructed from the APOGEE spectra. Any deviations could be due to observational or reduction issues in the light curves or they could result from information in the PSD that is not predictable from photospheric observations like stellar spectra. Examples of the latter are mode mixing or strong internal magnetic fields (e.g., \citealt{2015Sci...350..423F}).

In the near future, the all-sky \tess\ light-curves provide a great opportunity to expand this model especially with the two ecliptic continuous viewing zones. Proposed mission like PLATO and HAYDN \citep{2021ExA....51..963M} can provide even more accurate ages to use in the latent space age training set, as can more detailed asteroseismic modeling of individual modes in current data (e.g., \citealt{2021NatAs...5..640M}). More generally, advances in machine learning using artificial neural network allows more opportunities for training with cross-domain data with few available labels. In astronomy, it is very common to have observations across multiple-domain, for example in multi-messenger astronomy (e.g., the famous multi-messenger gravitational wave event GW170817; \citealt{2018LRR....21....3A}). Many of the objects on the sky are observed in multiple large-scale surveys; stellar spectra and light-curves as an example in this paper. The large amount of overlap between surveys in different domains can be exploited using methods such as the one used in this paper, because they contain more information than using one survey in one domain alone.

\begin{table*}
	\centering
	\caption{Data Model of \texttt{nn\_latent\_age\_dr17.csv} that is available here \url{https://github.com/henrysky/astroNN_ages/blob/main/nn\_latent\_age\_dr17.csv.gz} which is the data file generated from this paper, row-matched to the official \texttt{allStar-dr17-synspec\_rev1.fits} with $733,901$ rows in total while $308,119$ stars with latent representation computed with a subset of $142,257$ stars with latent space age measured. Missing data are represented by \texttt{NaN}. We recommend to use a latent space age uncertainty cut of $40\%$, stars without any \texttt{STARFLAG} and \texttt{ASPCAPFLAG} (see \secname\,\ref{subsec:realresult} and \secname\,\ref{sec:dr17_catalog} for discussion)}
	\begin{tabular}{lccl} % four columns, alignment for each
		\hline
		Label & Physical Units & Sources & Descriptions \\
		\hline
		\texttt{apogee\_id} & n/a & APOGEE DR17 & APOGEE ID \\
		\texttt{telescope} & n/a & APOGEE DR17 & Telescope used for APOGEE observation (\texttt{apo25m} or \text{lco25m}) \\
		\texttt{field} & n/a & APOGEE DR17 & APOGEE field name \\
  	\texttt{STARFLAG} & n/a & APOGEE DR17 & APOGEE star flag bitmask \\
		\texttt{ASPCAPFLAG} & n/a & APOGEE DR17 & APOGEE ASPCAP flag bitmask\\
		\texttt{z0} & n/a & This paper & $1^\mathrm{st}$ dimension of latent representation \\
		\texttt{z1} & n/a & This paper & $2^\mathrm{nd}$ dimension of latent representation \\
		\texttt{z2} & n/a & This paper & $3^\mathrm{rd}$ dimension of latent representation \\
		\texttt{z3} & n/a & This paper & $4^\mathrm{th}$ dimension of latent representation \\
		\texttt{z4} & n/a & This paper & $5^\mathrm{th}$ dimension of latent representation \\
		\texttt{LogAge} & $\log_{10}{(\mathrm{Gyr})}$ & This paper & Logarithmic latent space age \\
		\texttt{LogAge\_Error} & $\log_{10}{(\mathrm{Gyr})}$ & This paper & Logarithmic latent space age uncertainty \\
		\texttt{Age} & Gyr & This paper & Latent space age \\
		\texttt{Age\_Error} & Gyr & This paper & Latent space age uncertainty \\
		\hline
	\end{tabular}
	\label{table:data_model}
\end{table*}

\section*{Acknowledgements}

We thank the anonymous referee for helpful and constructive comments. HL and JB acknowledge financial support from NSERC (funding reference number RGPIN-2020-04712) and an Ontario Early Researcher Award (ER16-12-061). AM acknowledges support from the ERC Consolidator Grant funding scheme (project ASTEROCHRONOMETRY, G.A. n. 772293)

Funding for the Sloan Digital Sky Survey IV has been provided by the Alfred P. Sloan Foundation, the U.S. Department of Energy Office of Science, and the Participating Institutions. SDSS-IV acknowledges support and resources from the Center for High Performance Computing at the University of Utah. The SDSS website is www.sdss.org.

%%%%%%%%%%%%%%%%%%%%%%%%%%%%%%%%%%%%%%%%%%%%%%%%%%
\section*{Data Availability}

The basic data described in \secname~\ref{sec:data} is publicly available, with some data available in the \texttt{astroNN} DR17 VAC available as part of SDSS-IV's DR17. All of the code underlying this article and neural network models are available in a \texttt{Github} repository at \url{https://github.com/henrysky/astroNN_ages}. Tutorials are provided in the \texttt{Github} repository on how to load and use the models trained for this paper, including predicting a corresponding PSD from an APOGEE spectrum, getting a latent representation of an APOGEE spectrum, sampling a new PSD from the latent space, and performing the age determination from the latent space.



\bibliographystyle{mnras}
\bibliography{mnras_template} % if your bibtex file is called example.bib

%\begin{thebibliography}{99}
%\end{thebibliography}

%%%%%%%%%%%%%%%%%%%%%%%%%%%%%%%%%%%%%%%%%%%%%%%%%%

%%%%%%%%%%%%%%%%% APPENDICES %%%%%%%%%%%%%%%%%%%%%

% \appendix

% \section{Some extra material}

% If you want to present additional material which would interrupt the flow of the main paper,
% it can be placed in an Appendix which appears after the list of references.

%%%%%%%%%%%%%%%%%%%%%%%%%%%%%%%%%%%%%%%%%%%%%%%%%%

% Don't change these lines
\bsp	% typesetting comment
\label{lastpage}
\end{document}